%% file: neurips_2026.tex
\documentclass{article}

\PassOptionsToPackage{numbers}{natbib}

 \usepackage[main, final]{neurips_2026}

\usepackage[utf8]{inputenc} 
\usepackage[T1]{fontenc}    
\usepackage{hyperref}       
\usepackage{url}            
\usepackage{booktabs}       
\usepackage{amsfonts}       
\usepackage{nicefrac}       
\usepackage{microtype}      
\usepackage{pifont}       
\usepackage{xcolor}       

\newcommand{\cmark}{\textcolor{green}{\ding{51}}} 
\newcommand{\xmark}{\textcolor{red}{\ding{55}}}   
\usepackage{amsmath}
\usepackage{graphicx}
\usepackage{amsthm}  
\usepackage{multirow}
\usepackage{colortbl}
\usepackage{makecell}
\usepackage{subcaption}
\usepackage{algorithm}
\usepackage{algorithmic}
\definecolor{taskcolor}{HTML}{E4E4E4}   
\definecolor{shadowcolor}{HTML}{F2F2F2} 
\newcommand{\best}[1]{\textbf{#1}}
\newcommand{\sbest}[1]{\underline{#1}}
\usepackage[many,breakable]{tcolorbox}

\definecolor{boxbg}{HTML}{FCFBF4}     
\definecolor{boxborder}{HTML}{757373} 

\newtcolorbox{researchquestion}{
    enhanced,
    colback=boxbg,             
    colframe=boxborder,        
    boxrule=1pt,               
    arc=4mm,                   
    auto outer arc,
    top=1mm, bottom=1mm,       
    left=2mm, right=4mm,       
    sidebyside,                
    sidebyside align=center,   
    sidebyside gap=5mm,        
    lefthand width=1cm,      
    segmentation style={dashed, draw=boxborder, line width=0.5pt}, 
    fontlower=\itshape         
}
\newtcolorbox{promptbox}[1]{
  colback=gray!5!white,      
  colframe=gray!60,          
  colbacktitle=gray!20,      
  coltitle=black,            
  fonttitle=\bfseries,
  title=#1,
  arc=2mm,
  boxrule=0.5pt,             
  left=3mm, right=3mm, top=2mm, bottom=2mm,
  breakable
}

\theoremstyle{plain}
\newtheorem{definition}{Definition}[subsection]
\title{Group Perspective Matters: Regulating Debate Relationships Can Mitigate Blind Conformity in Multi-Agent Debate}

\author{%
  \textbf{Hao Wu}\textsuperscript{1,2}, 
  \textbf{Shoucheng Song}\textsuperscript{1,2}, 
  \textbf{Chang Yao}\textsuperscript{1,2},
  \textbf{Huaiyu Wan}\textsuperscript{1,2}, 
  \textbf{Youfang Lin}\textsuperscript{1,2}, 
  \textbf{Kai Lv}\textsuperscript{1,2}\thanks{Corresponding author.} \\
  \textsuperscript{1}School of Computer Science \& Technology,
  Beijing Jiaotong University, Beijing, China \\
  \textsuperscript{2}Beijing Key Laboratory of Traffic Data Mining and
  Embodied Intelligence, Beijing, China \\
  \texttt{\{insis\_wh,insis\_songsc,yaochang,hywan,yflin,lvkai\}@bjtu.edu.cn}
}

\begin{document}

\maketitle

\begin{abstract}
Multi-Agent Debate (MAD) improves the reasoning performance of Large Language Models (LLMs) through multi-round interaction. However, LLMs in MAD are highly susceptible to blind conformity. Existing individual evaluation methods, typically based on confidence or perplexity, fail to reflect the correctness of reasoning and may even exacerbate blind conformity. To address this, we shift the perspective from individual evaluation to group interaction. 
We define mutual referencing among LLMs as \textbf{Debate Relationships} and recognize that regulating these relationships is the key to mitigating blind conformity. 
In this paper, we propose a novel framework for \textbf{D}ynamically r\textbf{E}gulating deb\textbf{A}te \textbf{R}elationships (DEAR) from the group perspective. At first, DEAR quantifies consensus and divergence as \textit{group evidence} to capture the debate state. 
Then, DEAR operates through three stages: 1) What: perceiving group consultation tendency and uncertainty; 2) Who: introducing a Selection RL-Agent to dynamically select reference peers; and 3) How: adopting a Behavior RL-Agent to adaptively adjust generation behaviors. 
Notably, we formulate the execution of the two RL-Agents as a sequential decision-making process, jointly optimizing via multi-agent reinforcement learning.
Extensive experiments demonstrate that DEAR achieves superior performance while significantly reducing token consumption.
\end{abstract}

\section{Introduction}
\label{intro}
In recent years, Multi-Agent Debate (MAD) has gained widespread attention as an effective paradigm for enhancing the reasoning capabilities of Large Language Models (LLMs) \cite{du2024improving,chan2023chateval,hu2025debate}. Compared to the independent reasoning of a single model, MAD effectively mitigates hallucinations \cite{liang2024encouraging,michael2023debate,sun2025towards} by organizing multiple LLMs to debate around a common problem. However, in MAD, LLMs are highly susceptible to \textbf{blind conformity}—the minority often abandons their independent reasoning to align with the majority \cite{choi2025empirical,wang2026debate,zhou2026epistemic}.

Current methods evaluate individual LLM reasoning' reliability using confidence \cite{sun2025cortexdebate,bai2024confidencecal,zhu2026demystifying,yang2026mad} or perplexity \cite{tian2026multi}. By restricting the propagation of uncertain reasoning, these approaches moderately improve debate quality \cite{bai2024confidencecal}. However, both confidence and perplexity suffer from inherent flaws.
For confidence-based methods, they prompt LLMs to output confidence scores alongside their responses. Yet LLMs exhibit ``overconfidence'' \cite{yang2024confidence}, frequently assigning high confidence to incorrect reasoning even with confidence calibration \cite{sun2025cortexdebate,zhu2026demystifying}.
For perplexity-based methods, they calculate the token-level negative log-probability of the reasoning to quantify LLMs' uncertainty. Yet perplexity measures textual fluency rather than reasoning correctness, meaning LLMs may easily generate incorrect responses with low perplexity \cite{velivckovic2026perplexity,hu2024can}.
Moreover, the above methods apply fixed thresholds to propagate only high-confidence or low-perplexity responses. However, such unreliable individual evaluation may amplify the influence of incorrect reasoning and exacerbate blind conformity \cite{prasad2025two}.

The limitations of evaluation methods motivate us to rethink the nature of blind conformity: rather than a single reasoning error, blind conformity is a group phenomenon driven by majority dominance.  
As this phenomenon emerges from group interactions, static individual evaluations are insufficient to capture the evolving group dynamics.
Therefore, to mitigate blind conformity, we need to shift our perspective from individual evaluation to group interactions.
In group interactions, we argue that debate is a dynamic process of viewpoint collision and integration. During this process, LLMs continuously update their reasoning by referencing peer opinions. We define the above mutual referencing as \textbf{Debate Relationship}. 
This relationship spans the entire group interaction and serves as the medium for viewpoint exchange. A beneficial debate relationship facilitates the reasonable adoption of peer responses, enabling LLMs to enhance their reasoning.
Therefore, effectively regulating the debate relationships at the group level offers a viable pathway to mitigate the phenomenon of blind conformity.


Building on the above insight, we propose a novel framework for \textbf{D}ynamically r\textbf{E}gulating deb\textbf{A}te \textbf{R}elationships (DEAR) from the group perspective. To capture the debate state after each round, DEAR initially quantifies the group consensus and divergence as \textit{group evidence}. Leveraging this evidence, DEAR regulates the debate relationships through three stages: perceiving consultation tendency and uncertainty (\textbf{What}), selecting reference peers (\textbf{Who}), and guiding generation behaviors (\textbf{How}).
Specifically, in the first stage, we employ Subjective Logic (SL) \cite{jsang2018subjective} to decouple the group evidence into consultation tendency and uncertainty among LLMs.
In the second stage, we introduce a Selection RL-Agent to learn an adaptive peer selection policy to determine which peers each LLM should reference.
Finally, based on the established debate relationships, a Behavior RL-Agent is adopted to adjust the generation behaviors of individual LLMs. 
Furthermore, we formulate the ordered execution of the above RL-Agents as a sequential decision-making process. To guarantee the coordinated optimization of these policies, we innovatively employ the Heterogeneous-Agent Proximal Policy Optimization (HAPPO) algorithm \cite{kuba2021trust} for end-to-end joint training.
Extensive experiments on four mathematical reasoning tasks and four question-answering (QA) tasks demonstrate the effectiveness of our proposed method.

Our main contributions are summarized as follows:
\begin{itemize}
    \item We shift the perspective from individual evaluation to group interactions centered on debate relationships, effectively mitigating blind conformity in Multi-Agent Debate.
    
    \item We propose DEAR, a novel framework that employs group evidence to dynamically regulate debate relationships through: What–Who–How stages.
    
    
    \item We conduct extensive experiments to demonstrate that DEAR achieves superior performance while significantly reducing token consumption.
\end{itemize}

\section{Related Works}
\subsection{Single-Agent Reasoning.}
Several works propose various single-agent reasoning policies to augment the reasoning capabilities of LLMs \cite{wei2022chain,wang2022self,yao2023tree,besta2024graph,ziqi2023tab,shum2023automatic}. However, constrained by the intrinsic knowledge boundaries of an individual model, these methods struggle to correct hallucinations and reasoning errors.
\subsection{Multi-Agent Debate.}
Multi-Agent Debate (MAD) leverages multi-round LLM interactions to improve reasoning performance \cite{du2024improving,chan2023chateval,li2025advancing,yi2025debate}, yet LLMs frequently exhibit blind conformity during debate \cite{choi2025empirical,wang2026debate}. Existing methods filter responses using confidence or perplexity to improve debate quality \cite{sun2025cortexdebate,bai2024confidencecal,tian2026multi,yang2026mad}, but the inherent unreliability of these metrics may instead exacerbate blind conformity \cite{prasad2025two}. Free-MAD \cite{cui2025free} attempts to address the error propagation caused by incorrect majority responses through heuristic scoring of LLMs' reasoning trajectories. However, evaluating the correctness of individual reasoning is fundamentally infeasible during the debate process without ground truth.

More detailed related works and discussion can be seen in Appendix \ref{related_works}.


\section{Problem Definition}
\label{preliminary}
We aim to mitigate blind conformity in the Multi-Agent Debate (MAD) system by regulating debate relationships from the group perspective, without fine-tuning any internal parameters of Large Language Models (LLMs). In this setting, we formulate this regulation process as a Partially Observable Markov Decision Process (POMDP) defined by the tuple $\langle \mathcal{N}, \mathcal{V}, \mathcal{S}, \Omega, \mathcal{O}, \mathcal{I}, \mathcal{A}, \mathcal{T}, \mathcal{R}, \gamma \rangle$, where $\mathcal{N} = \{1, 2\}$ represents the set of the Selection RL-Agent (with policy $\rho$) and the Behavior RL-Agent (with policy $\pi$); $\mathcal{V} = \{1, \dots, V\}$ denotes the set of debating LLMs, with $V$ representing their number; $\mathcal{S}$ denotes the state space; $\Omega$ represents the observation space; $\mathcal{O}$ is the observation function; $\mathcal{I}$ and $\mathcal{A}$ denote the respective action spaces for the Selection and Behavior RL-Agents; $\mathcal{T}$ is the state transition function; $\mathcal{R}$ is the reward function that assigns an environmental reward $R_{debate}$ based on the correctness of the final answer at the terminal round $T$; and $\gamma$ is the discount factor.

Within this formulation, the Selection RL-Agent receives an observation $o^\rho \in \Omega$ derived from the group evidence, and outputs selection actions $\boldsymbol{ids} \in \mathcal{I}$ to regulate which peer opinions each LLM should reference.
The Behavior RL-Agent takes the observation and action of the Selection RL-Agent as conditional inputs, i.e., $o^\pi = (o^\rho, \boldsymbol{ids}) \in \Omega \times \mathcal{I}$, and outputs generation behaviors $\boldsymbol{a} \in \mathcal{A}$, where $a_i = [T_i, Top\_p_i]$, to adjust the temperature and nucleus sampling parameters for each LLM.
According to the probability chain rule, the joint policy of the two RL-Agents can be defined as:
\begin{equation}
\boldsymbol{\pi^{all}}(\boldsymbol{ids}, \boldsymbol{a} \mid s) = \rho(\boldsymbol{ids} \mid o^\rho) \cdot \pi(\boldsymbol{a} \mid o^\pi).
\end{equation}

\section{Method}

\begin{figure}[t]
    \centering
    \includegraphics[width=1.0\textwidth]{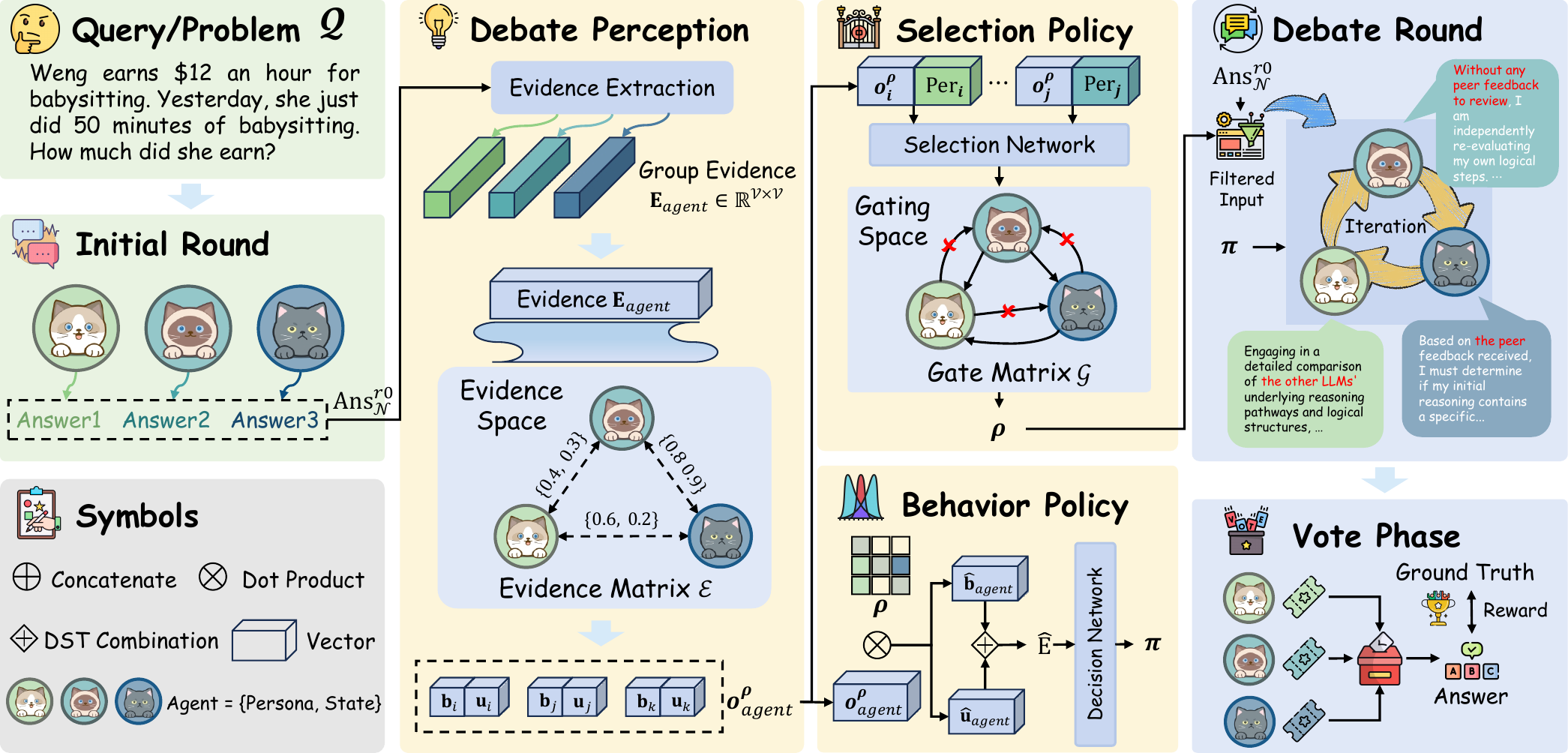} 
    \caption{
    The overall framework of DEAR. We employ a two-round debate as an example to detail our method. (1) \textit{Debate Perception}, consisting of extracting group evidence and \textbf{What}: perceiving group consultation tendency and uncertainty. (2) \textit{Selection Policy}, representing \textbf{Who}: adaptively selecting reference peers. (3) \textit{Behavior Policy}, denoting \textbf{How}: adjusting generation behavior.}
    \label{fig:framework}
\end{figure}

To mitigate blind conformity in MAD, our core idea is to dynamically regulate debate relationships from the group perspective. As illustrated in Figure \ref{fig:framework}, after each generation round, DEAR first quantifies the group consensus and divergence as group evidence to capture the debate state ($\triangleright$ Section \ref{sec:evidence}). Driven by this, DEAR operates through three stages:
(1) \textbf{What} ($\triangleright$ Section \ref{sec:what}): We apply Subjective Logic (SL) \cite{jsang2018subjective} to decouple the group evidence into consultation tendency and uncertainty, providing the foundation for regulating debate relationships.
(2) \textbf{Who} ($\triangleright$ Section \ref{sec:who}): A Selection RL-Agent is introduced to select which peer responses each LLM should reference, dynamically regulating the debate relationships.
(3) \textbf{How} ($\triangleright$ Section \ref{sec:how}): A Behavior RL-Agent is adopted to adjust the generation behaviors of LLMs within the established debate relationships.
Finally, we employ the Heterogeneous-Agent Proximal Policy Optimization (HAPPO) algorithm \cite{kuba2021trust} for joint training of the above RL-Agents ($\triangleright$ Section \ref{sec:optimizating}).



\subsection{Group Evidence Extraction}
\label{sec:evidence}

To dynamically regulate debate relationships, we should first realize the current debate state. During the debate, the consensus and divergence among LLMs directly represent the collision and integration of viewpoints, objectively reflecting the current group dynamics. Therefore, we quantify the above two critical indicators as group evidence to capture the debate state.

Specifically, we extract group evidence from paired LLM responses ($Ans_i$, $Ans_j$). Inherently, the final answers serve as the most direct indication of LLMs' viewpoint. The alignment between two answers directly signifies whether they are in consensus or in divergence. However, standard outputs (e.g., a number or option) yield excessively sparse semantics for meaningful comparison. To address this, we prompt each LLM to summarize its reasoning process into a conclusion. Subsequently, we quantify the consensus and divergence across two dimensions.

\paragraph{Conclusion Similarity.} 
Conclusions represent the LLMs' most direct judgments. To quantify conclusion similarity, we project the conclusions into a latent space $h_i \in \mathbb{R}^d$ using a pre-trained text encoder $\xi(\cdot)$ and calculate their scaled cosine similarity:
\begin{equation}
e_{ij}^{(con)} = \frac{1}{2}\left(1 + \frac{h_i^\top h_j}{\|h_i\|_2 \|h_j\|_2}\right).
\end{equation}
\paragraph{Reasoning Similarity.}
Conclusion similarity only captures the alignment of final judgments, but fails to reveal the underlying logical reasoning process. To address this, we further quantify reasoning similarity. Following the same procedure as conclusion similarity, we map the reasoning processes of LLMs into latent representations $c_i \in \mathbb{R}^d$. We then quantify the reasoning similarity with scaled cosine similarity:
\begin{equation}
e_{ij}^{(rea)} = \frac{1}{2}\left(1 + \frac{c_i^\top c_j}{\|c_i|_2 \|c_j|_2}\right).
\end{equation}


The above two similarities satisfy $e_{ij}^{(con)}, e_{ij}^{(rea)} \in [0,1]$. We adopt the arithmetic mean to compute the scalar evidence score $e_{ij}$ (where we set $e_{ii} = 0$). To capture the debate state, we aggregate all pairwise scores to formulate the overall group evidence matrix $\mathbf{E}_{agent}$. By measuring inter-LLM similarity independent of task-specific semantics, this evidence matrix is inherently task-agnostic and generalizes well across diverse reasoning tasks.

\subsection{Group Evidence Decouple (What)}
\label{sec:what}
While group evidence objectively captures consensus and divergence, it fails to reflect an LLM’s subjective consultation tendency toward its peers. Moreover, when facing conflicting opinions, LLMs inevitably experience uncertainty regarding peer credibility. This uncertainty is the key to perceiving whether they will abandon independent reasoning and blindly conform. Therefore, perceiving both the consultation tendency and the uncertainty is essential to effectively regulate debate relationships.

To realize this, we employ Subjective Logic (SL) \cite{jsang2018subjective}. SL provides a theoretical framework that translates evidence into a distribution of belief masses across multiple targets, accompanied by an overall uncertainty mass. Accordingly, we decouple LLM $i$'s evidence into two components: the consultation tendency (the belief mass) $b_{ij}$ toward peer $j$, and the corresponding uncertainty $u_i$. These variables satisfy the following zero-sum constraint:
\begin{equation}
u_i + \sum_{j \in \mathcal{V}} b_{ij} = 1, \quad b_{ij} \ge 0, \quad u_i \ge 0.
\end{equation}

In SL, transforming evidence into the belief mass requires a base weight to represent the initial total uncertainty (i.e., when no evidence has been collected). We set this base weight to the number of LLMs $V$. Defining the total accumulated evidence as $E_i = \sum_{j \in \mathcal{V}} e_{ij}$, the overall evidence strength of LLM $i$ is $S_i = E_i + V$. The belief mass $b_{ij}$ and the uncertainty $u_i$ can be derived as:
\begin{equation}
b_{ij} = \frac{e_{ij}}{S_i}, \quad u_i = \frac{V}{S_i}.
\end{equation}

Based on this formulation, we perceive the LLMs' subjective consultation tendencies and uncertainties to serve the subsequent regulation.

\subsection{Adaptive Debater Selection (Who)}
\label{sec:who}
With the consultation tendencies and uncertainties, we adopt them to regulate the debate relationships by determining the specific peers each LLM should reference. However, relying on static heuristic rules to make decisions is inadequate for the evolving dynamics of debate. Thus, we introduce a Selection RL-Agent to learn an adaptive peer selection policy $\rho$ for group-level regulation.

To achieve this, we construct a group observation $o^{\rho}$ by concatenating the consultation tendency $b_i$, uncertainty $u_i$, and a one-hot identity embedding $ID_i \in \{0,1\}$ of all LLMs:
\begin{equation}
o^{\rho} = \left[ b_1 \parallel u_1 \parallel \mathrm{ID_1} \parallel \dots \parallel b_V \parallel u_V \parallel \mathrm{ID_V} \right].
\end{equation}

Then, we learn a mapping network $f_{\rho}: \mathbb{R}^{V(2V+1)} \rightarrow [0,1]^{V \times (V-1)}$ that maps $o^\rho$ to a preference matrix $\mathcal{P} = [p_{i \leftarrow j}]_{i \neq j}$.
Fundamentally, regulating debate relationships is a binary decision-making process: an LLM either references a peer's response or does not. To realize this, we formulate the discrete peer selection action $\boldsymbol{ids} \in \mathcal{I}$ as a set of independent Bernoulli variables parameterized by $\mathcal{P}$. Accordingly, each binary action $ids_{i \leftarrow j} \in \{0, 1\}$, indicating whether LLM $i$ references peer $j$, is sampled from the following joint policy distribution:
\begin{equation}
\boldsymbol{ids} \sim \rho(\boldsymbol{ids} \mid o^\rho) = \prod_{i=1}^{V} \prod_{\substack{j=1, \ j \neq i}}^{V} (p_{i \leftarrow j})^{ids_{i \leftarrow j}} (1 - p_{i \leftarrow j})^{1 - ids_{i \leftarrow j}}.
\end{equation}

The action $\boldsymbol{ids}$ serves as the interaction mask for the subsequent debate round. 
By achieving group-level regulation of debate relationships, the RL-Agent selectively severs the transmission of harmful responses while preserving beneficial reasoning exchange, thereby mitigating blind conformity.

\subsection{Generation Behavior Adjustment (How)}
\label{sec:how}
After selecting a distinct subset of reference peers, each LLM should integrate the selected peers' evidence into its own decisions. Traditional mean aggregation fails to distinguish between consensus and divergence, which are essential for determining whether an LLM should converge on existing opinions or explore breakthroughs.
Instead, we employ the Dempster-Shafer Evidence Theory (DST) combination rule to fuse evidence.

\begin{definition}[\textit{\textbf{Dempster's Combination Rule}}]
Given two independent belief sets $\mathcal{M}_1 = (\{b_{1k}\}_{k=1}^{V}, u_1)$ and $\mathcal{M}_2 = (\{b_{2k}\}_{k=1}^{V}, u_2)$, their fused joint belief $\mathcal{M} = \mathcal{M}_1 \oplus \mathcal{M}_2 = (\{b_k\}_{k=1}^{V}, u)$ is computed as:
$$b_k = \frac{1}{1-C} (b_{1k}b_{2k} + b_{1k}u_2 + b_{2k}u_1),\quad u = \frac{1}{1-C} u_1 u_2,$$
where $C = \sum_{p \neq q} b_{1p}b_{2q}$ quantifies the degree of conflict between the two beliefs, and $\frac{1}{1-C}$ is the normalization factor. 
\end{definition}

Based on this, LLM $i$ fuses its own belief set $\mathcal{M}_i$ with peers from the selected subset $\mathcal{X}_i = \{j \in \mathcal{V} \mid j \neq i, ids_{i \leftarrow j} = 1\}$ by the subsequent rule:
\begin{equation}
{\mathcal{M}} = \mathcal{M}_1 \oplus \mathcal{M}_2 \oplus \dots \mathcal{M}_V.
\end{equation}
Once we have determined the fused belief $\hat{\mathcal{M}}_i = \left( \hat{b}_i, \hat{u}_i \right)$, the evidence $\hat{E}_i$ can be derived as: $\hat{S}_i = \frac{V}{\hat{u}_i}, \hat{E}_{i} = \hat{b}_{i} \times \hat{S}_i$.
Taking $\hat{E}_i$ as input, we sample the continuous variable $z_i = [z_{i, T}, z_{i, P}]^\top$ from a diagonal Gaussian distribution parameterized by the Behavior RL-Agent policy $\pi$:
\begin{equation}
z_i \sim \pi(\cdot \mid \hat{E}_i) = \mathcal{N}\left(\boldsymbol{\mu}(\hat{E}_i), \text{diag}(\boldsymbol{\nu}(\hat{E}_i)^2)\right).
\end{equation}
To ensure the generation behaviors fall within the valid ranges $[T_{min}, T_{max}]$ and $[P_{min}, P_{max}]$, we apply a scaling mapping to the sampled variable $z_i$:
\begin{equation}
T_i = T_{min} + (T_{max} - T_{min}) \cdot \sigma(z_{i, T}),
\end{equation}
\begin{equation}
Top\_p_i = P_{min} + (P_{max} - P_{min}) \cdot \sigma(z_{i, P}),
\end{equation}
where $\sigma(\cdot)$ denotes the Sigmoid function. 
Finally, the Behavior RL-Agent outputs the continuous generation behavior $a_i = [T_i, Top\_p_i]$, effectively adjusting the LLM's generation behavior in the subsequent debate round.

\subsection{Optimization Scheme}
\label{sec:optimizating}

Within DEAR, the Selection RL-Agent first generates the peer selection action $\boldsymbol{ids}$. Subsequently, the Behavior RL-Agent dynamically adjusts the LLM's generation behavior based on $\boldsymbol{ids}$. This ordered execution constitutes an explicit sequential decision-making process. To ensure coordinated optimization, we employ the Heterogeneous-Agent Proximal Policy Optimization (HAPPO) algorithm \cite{kuba2021trust} for end-to-end joint training.

\paragraph{The Selection RL-Agent Optimization.} 
Given that the Selection RL-Agent acts first in each iteration, we prioritize its optimization. Taking $o^{\rho}$ as its input, this RL-Agent outputs a binary action $\boldsymbol{ids}$ for all LLMs. Its policy update ratio is defined as $r_\phi^{\rho, t} = \frac{\rho_{\phi}(\boldsymbol{ids^t} \mid o^{\rho})}{\rho_{\phi_{old}}(\boldsymbol{ids^t} \mid o^{\rho})}$. The Selection RL-Agent is optimized based on the advantage function $\hat{A}^t$, with the objective formulated as:
\begin{equation}
\label{loss:selection}
\mathcal{L}_{\rho}(\phi) = \mathbb{E} \left[ - \min \left( r_\phi^{\rho, t} \hat{A}^t, \text{clip}(r_\phi^{\rho, t}, 1\pm\epsilon) \hat{A}^t \right) \right].
\end{equation}
Here, $\hat{A}^t$ represents the Generalized Advantage Estimation (GAE), which is computed via the Temporal Difference (TD) error $\delta^t$:
\begin{equation}
\hat{A}^{t}=\sum_{l=0}^{T-t}(\gamma\lambda)^{l}\delta^{t+l}, \quad
\delta^{t}=\mathcal{R}^{t}+\gamma V_{\psi}(s^{t+1})-V_{\psi}(s^{t}), \quad
\mathcal{R}^t = \begin{cases} 
R_{debate}, & t = T \\ 
0, & t < T 
\end{cases},
\end{equation}
where $R_{debate}$ denotes the environmental reward, $T$ denotes the number of debate rounds.

\paragraph{The Behavior RL-Agent Optimization.} 
Following the update of the Selection RL-Agent, we proceed to optimize the Behavior RL-Agent. Specifically, the sequential surrogate advantage for the Behavior RL-Agent is defined as: $M^{\pi, t} = r_\phi^{\rho, t} \hat{A}^t,$ where $r_\phi^{\rho, t}$ is the importance sampling ratio previously computed by the Selection RL-Agent. 
This term accounts for the distributional change induced by the
updated Selection policy when optimizing the conditional Behavior policy.
Since the Sigmoid mapping is a deterministic transformation, the essence of updating the Behavior RL-Agent is to optimize the parameterized distribution of the variable $\boldsymbol{z}^t$. 
Therefore, the corresponding policy update ratio is computed as $r_\theta^{\pi, t} = \frac{\pi_{\theta}(\boldsymbol{z}^t \mid \hat{E}^t)}{\pi_{\theta_{old}}(\boldsymbol{z}^t \mid \hat{E}^t)}$. The optimization objective is defined as:
\begin{equation}
\label{loss:behavior}
\mathcal{L}_{\pi}(\theta) = \mathbb{E} \left[ - \min \left( r_\theta^{\pi, t} M^{\pi, t}, \text{clip}(r_\theta^{\pi, t}, 1\pm\epsilon) M^{\pi, t} \right) \right].
\end{equation}
\paragraph{The Global Critic Optimization.}
To support the advantage estimation for the above RL-Agents, the centralized critic $V_\psi(s)$ precisely fits the expected return $\hat{r}^t = \hat{A}^t + V_{\psi_{old}}(s^t)$. We additionally apply value clipping to stabilize critic updates. The objective is expressed as:
\begin{equation}
\label{loss:critic}
\mathcal{L}_{critic}(\psi) = \mathbb{E} \left[ \max \left( (V_\psi(s^t) - \hat{r}^t)^2, \left(\text{clip}(V_\psi(s^t), V_{\psi_{old}}(s^t) \pm \epsilon) - \hat{r}^t\right)^2 \right) \right].
\end{equation}

The pseudo-code for DEAR is given in Appendix \ref{pseudocode}. Through the end-to-end training, we obtain the coordinated pair of Selection RL-Agent and Behavior RL-Agent for subsequent evaluation.

\begin{table*}[t]
\centering
\caption{Performance comparison on \texttt{Qwen3-8B}. We highlight the \textbf{optimal} and \underline{suboptimal} results. \textbf{Acc. (\%)} denotes accuracy, and \textbf{T. ($K$)} represents the average tokens per task.}
\label{tab:main_results_split}
\resizebox{\textwidth}{!}{%
\begin{tabular}{cc *{10}{c}}
\toprule

\rowcolor{taskcolor}
\multicolumn{12}{c}{\textbf{Math Reasoning Tasks}} \\

\multirow{3}{*}{\textbf{Architecture}}
& \multirow{3}{*}{\textbf{Method}}
& \multicolumn{2}{c}{\textbf{ID}}
& \multicolumn{6}{c}{\textbf{OOD}}
& \multicolumn{2}{c}{\multirow{2}{*}{\textbf{Avg. (Math)}}} \\
\cmidrule(lr){3-4}
\cmidrule(lr){5-10}

& & \multicolumn{2}{c}{\textbf{GSM8K}}
  & \multicolumn{2}{c}{\textbf{AIME24}}
  & \multicolumn{2}{c}{\textbf{GSM-Hard}}
  & \multicolumn{2}{c}{\textbf{MATH-500}}
  & \multicolumn{2}{c}{} \\

& & Acc. $\uparrow$ & T. $\downarrow$
  & Acc. $\uparrow$ & T. $\downarrow$
  & Acc. $\uparrow$ & T. $\downarrow$
  & Acc. $\uparrow$ & T. $\downarrow$
  & Acc. $\uparrow$ & T. $\downarrow$ \\
\midrule

\multirow{2}{*}{Single-Agent}
& CoT
& 91.5 & 0.4
& 30.0 & 4.1
& 57.0 & 0.4
& 70.8 & 1.2
& 62.3 & 1.5 \\

& CoT-SC
& \sbest{93.5} & 5.4
& \best{36.7} & 64.2
& 60.3 & 8.4
& 75.8 & 16.9
& \sbest{66.6} & 23.7 \\
\midrule

\multirow{7}{*}{Multi-Agent}
& Majority Voting
& 93.0 & 1.1
& \sbest{33.3} & 11.7
& 60.5 & 1.7
& 74.0 & 3.4
& 65.2 & 4.5 \\

& MAD
& \sbest{93.5} & 11.1
& 30.0 & 75.7
& 61.5 & 16.3
& 76.3 & 30.8
& 65.3 & 33.5 \\

& DMAD
& 91.5 & 17.5
& \sbest{33.3} & 98.2
& \sbest{61.8} & 23.4
& \sbest{78.8} & 41.0
& 66.4 & 45.0 \\

& CortexDebate
& \best{94.0} & 15.9
& 30.0 & 100.0
& 60.5 & 21.3
& 69.3 & 35.6
& 63.5 & 43.2 \\

& MAD-M$^2$(O)
& 92.0 & 8.7
& 30.0 & 47.9
& 61.5 & 10.0
& 77.0 & 20.8
& 65.1 & 21.9 \\

& ECON
& 91.0 & 8.0
& -- & --
& 61.0 & 9.3
& 72.0 & 13.3
& -- & -- \\

\rowcolor{shadowcolor}
\cellcolor{white}
& DEAR (Ours)
& \best{94.0} & 7.6
& \best{36.7} & 21.2
& \best{73.5} & 6.2
& \best{85.0} & 8.4
& \best{72.3} & 10.9 \\

\midrule

\rowcolor{taskcolor}
\multicolumn{12}{c}{\textbf{QA Tasks}} \\

\multirow{3}{*}{\textbf{Architecture}}
& \multirow{3}{*}{\textbf{Method}}
& \multicolumn{2}{c}{\textbf{ID}}
& \multicolumn{6}{c}{\textbf{OOD}}
& \multicolumn{2}{c}{\multirow{2}{*}{\textbf{Avg. (QA)}}} \\
\cmidrule(lr){3-4}
\cmidrule(lr){5-10}

& & \multicolumn{2}{c}{\textbf{ARC-C}}
  & \multicolumn{2}{c}{\textbf{MMLU (Pro.H.)}}
  & \multicolumn{2}{c}{\textbf{TruthfulQA}}
  & \multicolumn{2}{c}{\textbf{GPQA Dia.}}
  & \multicolumn{2}{c}{} \\

& & Acc. $\uparrow$ & T. $\downarrow$
  & Acc. $\uparrow$ & T. $\downarrow$
  & Acc. $\uparrow$ & T. $\downarrow$
  & Acc. $\uparrow$ & T. $\downarrow$
  & Acc. $\uparrow$ & T. $\downarrow$ \\
\midrule

\multirow{2}{*}{Single-Agent}
& CoT
& 90.0 & 0.4
& 56.8 & 0.4
& 63.3 & 0.5
& 36.7 & 2.0
& 61.7 & 0.8 \\

& CoT-SC
& \best{95.0} & 8.1
& 61.5 & 6.2
& 70.5 & 7.5
& 40.7 & 25.6
& 66.9 & 11.9 \\
\midrule

\multirow{7}{*}{Multi-Agent}
& Majority Voting
& 91.0 & 1.5
& \sbest{66.5} & 2.1
& \sbest{74.0} & 1.9
& 41.3 & 6.5
& \sbest{68.2} & 3.0 \\

& MAD
& 90.0 & 15.7
& 62.5 & 30.8
& 68.8 & 20.2
& 40.7 & 30.4
& 65.5 & 24.3 \\

& DMAD
& 91.5 & 22.6
& 64.0 & 39.2
& 72.0 & 27.7
& 38.0 & 64.1
& 66.4 & 38.4 \\

& CortexDebate
& 91.8 & 11.8
& 60.0 & 19.8
& 70.8 & 12.6
& \sbest{46.0} & 61.8
& 67.2 & 26.5 \\

& MAD-M$^2$(O)
& 90.0 & 4.7
& 65.0 & 14.5
& 73.0 & 6.8
& 42.7 & 29.8
& 67.7 & 14.0 \\

& ECON
& 90.5 & 13.3
& 64.8 & 10.9
& 64.5 & 9.2
& 31.3 & 16.4
& 62.8 & 12.5 \\

\rowcolor{shadowcolor}
\cellcolor{white}
& DEAR (Ours)
& \sbest{93.3} & 4.6
& \best{68.5} & 6.9
& \best{80.8} & 4.8
& \best{52.0} & 13.9
& \best{73.7} & 7.6 \\

\bottomrule
\end{tabular}%
}
\end{table*}

\section{Experiments}
In this section, we design extensive experiments to answer: (\textbf{RQ1}) How does DEAR perform against existing MAD methods in terms of both accuracy and token consumption?
(\textbf{RQ2}) How scalable is DEAR across varying foundation models and debate configurations?
(\textbf{RQ3}) What is the specific contribution of key components to the performance of DEAR?

\subsection{Experimental Settings}
\paragraph{Datasets and Benchmarks.}
To evaluate the effectiveness of DEAR, we employ four math reasoning datasets (i.e., GSM8K \cite{cobbe2021training}, AIME24, GSM-Hard \cite{gao2022pal}, and MATH-500 \cite{lightman2023lets}) and four question answering (QA) datasets (i.e., ARC-C \cite{allenai:arc}, MMLU\_Pro (Health) \cite{wang2024mmlu}, TruthfulQA \cite{lin2022truthfulqa}, and GPQA Diamond \cite{rein2024gpqa}). Among these, AIME24 and GPQA Diamond are challenging tasks.

\paragraph{Baselines.}
For single-agent baselines, we adopt Chain-of-Thought (CoT) \cite{wei2022chain} and Self-Consistency Chain-of-Thoughts (CoT-SC) \cite{wang2022self} with 16 independent reasoning paths. For multi-agent baselines, we employ Majority Voting, Multi-Agent Debate (MAD) \cite{du2024improving}, Diverse Multi-Agent Debate (DMAD) \cite{liu2025breaking}, and CortexDebate \cite{sun2025cortexdebate}. We employ MAD-M²(O) \cite{tian2026multi} for open-source LLMs, and regarding the MAD-M$^2$(O) for perplexity calculation, we adopt its variant, MAD-M$^2$(S) \cite{tian2026multi} for closed-source LLMs. Additionally, for training-based reinforcement learning methods, we select ECON \cite{yi2025debate} as our baseline. 
Notably, we refrain from fine-tuning any internal parameters of LLMs and exclusively train the Selection RL-Agent and the Behavior RL-Agent.

Detailed descriptions of the benchmarks and baselines are provided in Appendix \ref{dataset and baseline}.

\paragraph{Implementation Details.}
Only divergent viewpoints are worthy of debate. 
The diverse responses of LLMs can introduce multidimensional perspectives into MAD \cite{yang2026understanding,nguyen2026hear,zhu2026demystifying,li2026dynadebate}. Therefore, for the homogeneous debate system, we assign a distinct persona to each LLM following \cite{yang2026understanding}. To ensure fair comparison, all baselines are initialized with the same personas as DEAR. For single-agent baselines, we compute the average accuracy across different personas. 
The pre-trained text encoder $\xi(\cdot)$ uses BGE-M3 \cite{bge-m3}. 
Comprehensive LLM persona configurations and additional implementation details are provided in Appendix \ref{implementation detail}.

\subsection{Main Results \textbf{(RQ1)}}
We evaluate the performance of DEAR using both an open-source model (\texttt{Qwen3-8B}) \cite{qwen3technicalreport} and a closed-source model (\texttt{GPT-4o-mini}). 
For math reasoning tasks, DEAR is trained on the GSM8K training set and evaluated on all math test sets, with GSM8K as the in-distribution (ID) benchmark and the others as out-of-distribution (OOD) benchmarks. For QA tasks, DEAR is trained on the ARC-C training set and evaluated on all QA test sets, with ARC-C as ID and the others as OOD.
Table \ref{tab:main_results_split} presents the evaluation results, with additional results provided in Appendix Table \ref{tab:main_results_split_gpt}. Our key observations are summarized as follows:


\textbf{\textit{Observation 1}: DEAR achieves the optimal or suboptimal results across diverse task domains}. As shown in Table \ref{tab:main_results_split} and \ref{tab:main_results_split_gpt}, DEAR significantly outperforms baseline methods in average accuracy for both math reasoning and QA tasks. 
Even in challenging datasets, AIME24 and GPQA Diamond, DEAR yields substantial performance gains. 
These results show that 
MAD’s fully-connected topology tends to amplify erroneous reasoning. Meanwhile, 
individual evaluation methods such as CortexDebate and MAD-M$^2$(O) rely on unreliable confidence or perplexity, which exacerbate blind conformity and compromise overall performance. 
In contrast, DEAR regulates debate relationships based on objective group evidence rather than flawed metrics. Using this evidence, the Selection RL-Agent dynamically selects which peers to reference to shield reasonable minority reasoning. Meanwhile, the Behavior RL-Agent utilizes DST-fused evidence to adaptively balance exploration and convergence, guiding the LLMs toward a valid consensus.

\textbf{\textit{Observation 2}: DEAR significantly reduces token consumption during the debate}. Current methods typically rely on massive token expenditure to improve accuracy. This reliance is especially obvious in debate methods such as DMAD and individual LLM policies like CoT-SC. In contrast, DEAR's efficiency stems from the dynamic peer selection driven by the Selection RL-Agent. By perceiving the consultation tendency and uncertainty, the RL-Agent adaptively severs valueless peer references, preserving the beneficial reasoning exchanges, thereby reducing token consumption.

\begin{table*}[t]
\centering
\caption{Performance and token consumption comparison on advanced closed-source models.}
\label{tab:cross_model_results}
\resizebox{\textwidth}{!}{%
\begin{tabular}{cc *{16}{c}}
\toprule

\rowcolor{taskcolor}
\multicolumn{2}{c}{}
& \multicolumn{8}{c}{\textbf{Math Reasoning Tasks}}
& \multicolumn{8}{c}{\textbf{QA Tasks}} \\

\multirow{3}{*}{\textbf{LLM}}
& \multirow{3}{*}{\textbf{Method}}
& \multicolumn{2}{c}{\textbf{ID}}
& \multicolumn{4}{c}{\textbf{OOD}}
& \multicolumn{2}{c}{\multirow{2}{*}{\textbf{Avg. (Math)}}}
& \multicolumn{2}{c}{\textbf{ID}}
& \multicolumn{4}{c}{\textbf{OOD}}
& \multicolumn{2}{c}{\multirow{2}{*}{\textbf{Avg. (QA)}}} \\
\cmidrule(lr){3-4}
\cmidrule(lr){5-8}
\cmidrule(lr){11-12}
\cmidrule(lr){13-16}

& & \multicolumn{2}{c}{\textbf{GSM-Hard}}
  & \multicolumn{2}{c}{\textbf{AIME24}}
  & \multicolumn{2}{c}{\makecell{\textbf{IMO}\\\textbf{AnswerBench}}}
  & \multicolumn{2}{c}{}
  & \multicolumn{2}{c}{\textbf{TruthfulQA}}
  & \multicolumn{2}{c}{\makecell{\textbf{MedXpertQA}\\\textbf{(Text)}}}
  & \multicolumn{2}{c}{\textbf{GPQA Dia.}}
  & \multicolumn{2}{c}{} \\

& & Acc. $\uparrow$ & T. $\downarrow$
  & Acc. $\uparrow$ & T. $\downarrow$
  & Acc. $\uparrow$ & T. $\downarrow$
  & Acc. $\uparrow$ & T. $\downarrow$
  & Acc. $\uparrow$ & T. $\downarrow$
  & Acc. $\uparrow$ & T. $\downarrow$
  & Acc. $\uparrow$ & T. $\downarrow$
  & Acc. $\uparrow$ & T. $\downarrow$ \\
\midrule

\multirow{4}{*}{\makecell{\texttt{Claude-4.5-}\\\texttt{Sonnet}}}
& MAD
& 65.0 & 8.74
& 60.0 & 40.7
& 23.0 & 39.9
& 49.3 & 29.8
& 74.0 & 5.95
& \sbest{35.0} & 29.1
& 76.0 & 32.2
& 61.7 & 22.4 \\

& CortexDebate
& 69.0 & 12.64
& 53.3 & 57.6
& 21.0 & 57.3
& 47.8 & 42.5
& \sbest{79.0} & 7.35
& 29.0 & 48.4
& 74.0 & 50.4
& 60.7 & 35.4 \\

& MAD-M$^2$(S)
& \sbest{70.0} & 14.39
& \sbest{63.3} & 44.8
& \sbest{25.0} & 40.0
& \sbest{52.8} & 33.1
& 77.0 & 6.49
& 31.0 & 30.7
& \sbest{78.0} & 36.5
& \sbest{62.0} & 24.6 \\

\rowcolor{shadowcolor}
\cellcolor{white}
& DEAR (Ours)
& \best{76.0} & 7.26
& \best{76.7} & 26.6
& \best{35.0} & 31.7
& \best{62.6} & 21.9
& \best{83.0} & 4.27
& \best{40.0} & 23.2
& \best{82.0} & 24.0
& \best{68.3} & 17.2 \\

\midrule

\multirow{4}{*}{\makecell{\texttt{Gemini-3.1-}\\\texttt{Flash-Lite}}}
& MAD
& 68.0 & 15.93
& 63.3 & 35.4
& 22.0 & 34.1
& 51.1 & 28.5
& 89.0 & 13.75
& 54.0 & 25.3
& 73.0 & 28.0
& 72.0 & 22.4 \\

& CortexDebate
& 71.0 & 17.25
& \sbest{66.7} & 36.0
& 21.0 & 50.4
& 52.9 & 34.6
& \sbest{91.0} & 16.81
& 44.0 & 31.2
& 80.0 & 38.8
& 71.7 & 28.9 \\

& MAD-M$^2$(S)
& \sbest{72.0} & 19.63
& \sbest{66.7} & 33.1
& \sbest{24.0} & 33.3
& \sbest{54.2} & 28.7
& 90.0 & 18.45
& \sbest{55.0} & 23.2
& \sbest{81.0} & 27.9
& \sbest{75.3} & 23.2 \\

\rowcolor{shadowcolor}
\cellcolor{white}
& DEAR (Ours)
& \best{75.0} & 12.74
& \best{73.3} & 19.6
& \best{27.0} & 19.1
& \best{58.4} & 17.1
& \best{93.0} & 10.28
& \best{59.0} & 10.6
& \best{84.0} & 13.6
& \best{78.7} & 11.5 \\

\bottomrule
\end{tabular}%
}
\end{table*}

\begin{figure*}[t]
    \centering
    \includegraphics[width=1.0\textwidth]{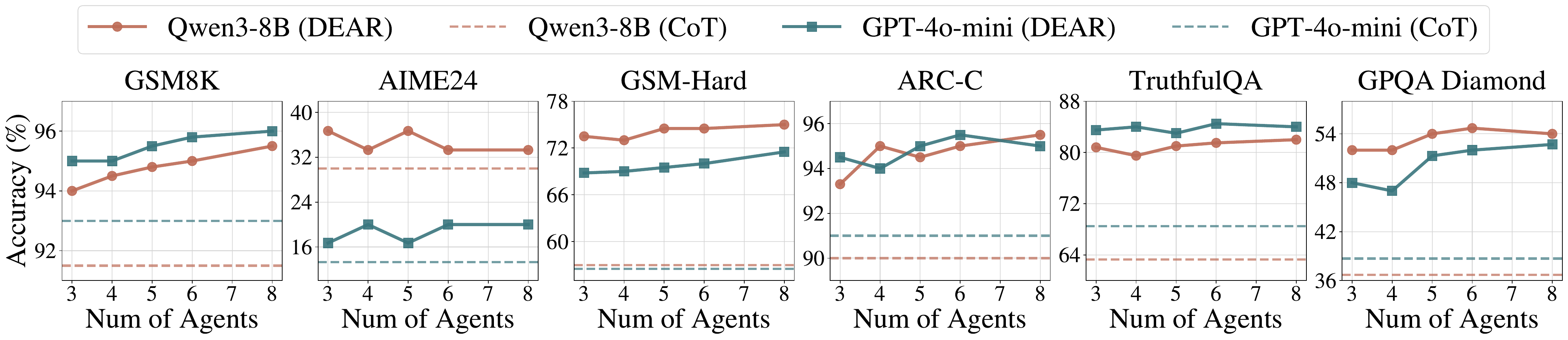} 
    \caption{Effect of scaling the number of agents in the case of \texttt{Qwen3-8B} and \texttt{GPT-4o-mini}.}
    \label{fig:agents}
\end{figure*}

\subsection{Scalability Analysis \textbf{(RQ2)}}
This section evaluates DEAR across two dimensions: (1) adaptability across open- and closed-source models, and (2) scalability with respect to the number of debate rounds and LLM agents.

\paragraph{Different Model Adaptability.}
To verify compatibility, we test DEAR on advanced closed-source models \texttt{Claude-4.5-Sonnet} and \texttt{Gemini-3.1-Flash-Lite}. 
We also incorporate the exceptionally challenging datasets IMOAnswerBench \citep{luong-etal-2025-towards} and MedXpertQA (Text) \citep{zuo2025medxpertqa}, while randomly sampling 200 examples from GSM-Hard and TruthfulQA to construct the math and QA training sets, respectively. As shown in Table \ref{tab:cross_model_results}, DEAR seamlessly adapts to diverse models while achieving superior performance on complex tasks.
Several existing MAD methods rely on extracting token-level probabilities to calculate perplexity \cite{qiao2026epistemic,tian2026multi}. However, closed-source models and APIs restrict access to these probability distributions, severely limiting their applicability. DEAR eliminates this reliance by utilizing group evidence, which quantifies inter-LLM similarity based solely on semantic outputs, thereby decoupling debate regulation from internal token distributions and specific tasks. 

\begin{figure*}[t]
    \centering
    \includegraphics[width=1.0\textwidth]{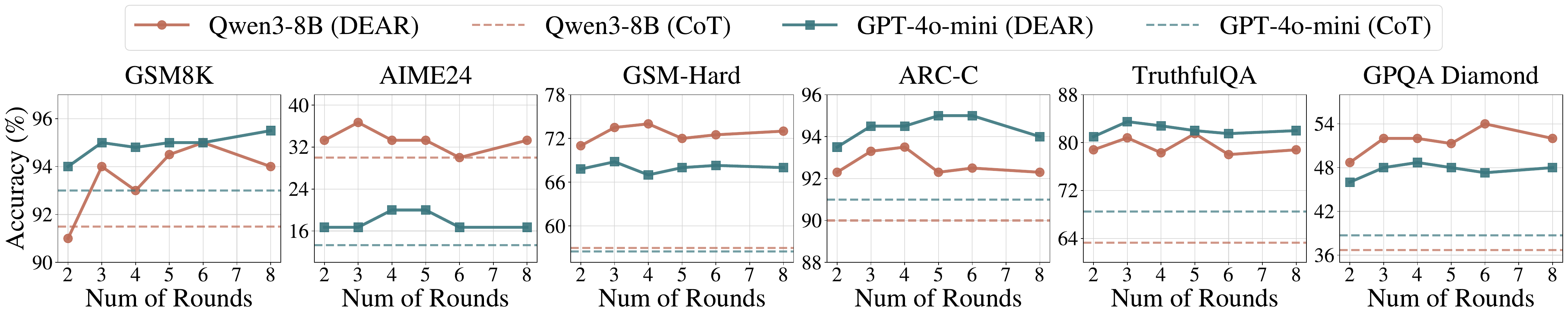} 
    \caption{Effect of scaling the number of debate rounds in the case of \texttt{Qwen3-8B} and \texttt{GPT-4o-mini}.}
    \label{fig:rounds}
\end{figure*}

\begin{figure*}[t]
    \centering
    \includegraphics[width=1.0\textwidth]{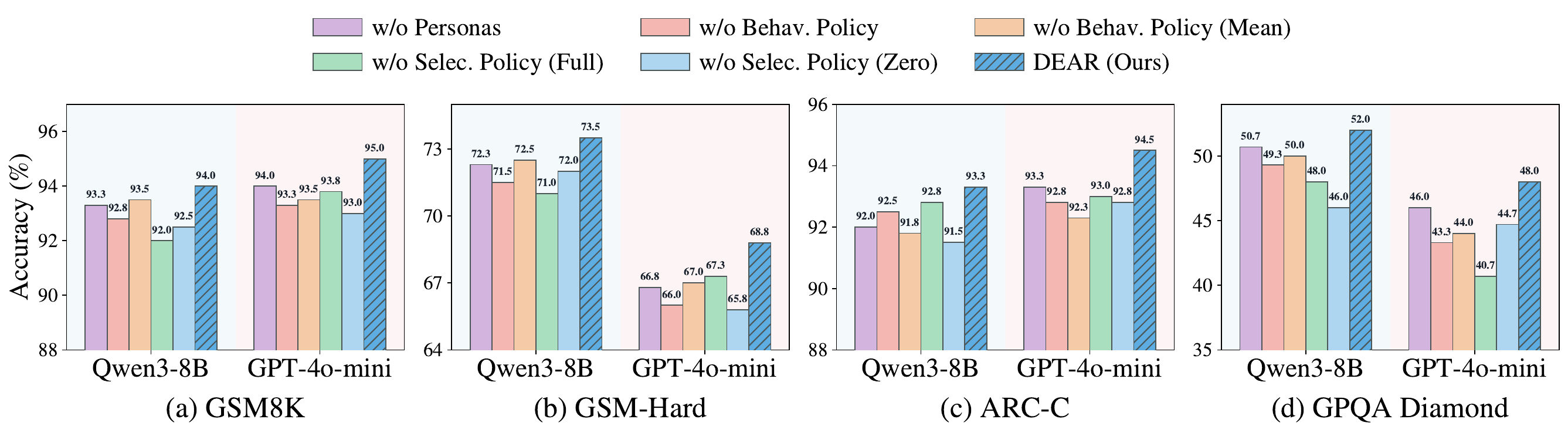} 
    \caption{Ablation study of the proposed DEAR on four datasets.}
    \label{fig:ablation}
\end{figure*}

\paragraph{Scaling of Debate Rounds \& Number of LLMs.}
In this section, we evaluate the scalability of our framework. When scaling the number of LLMs, we fix the debate at three rounds. We also assign specific personas to each LLM following the configuration in \cite{yang2026understanding}. Conversely, when scaling the debate rounds, we fix the LLM count at three. As illustrated in Figures \ref{fig:agents} and \ref{fig:rounds}, the performance of DEAR consistently improves as the number of LLMs increases. This demonstrates that the Selection RL-Agent effectively identifies reliable peers. In terms of scaling the debate rounds, DEAR converges to near-optimal performance within fewer rounds. This is attributed to the Behavior RL-Agent's adjustment of generation behaviors, which facilitates valid consensus formation in the early rounds.

\subsection{Ablation Study \textbf{(RQ3)}}
To validate the effectiveness of each components within DEAR, we design several ablated variants. Specifically, we remove diverse personas (w/o Personas). For the Behavior RL-Agent, we either completely remove it (w/o Behav. Policy) or substitute the DST fusion rule with simple mean-pooling (w/o Behav. Policy (Mean)). As for the Selection RL-Agent, we introduce two variants: a fully connected setting where all debaters are involved (w/o Selec. Policy (Full)), and a fully disconnected setting where selection is disabled (w/o Selec. Policy (Zero)).

Figure \ref{fig:ablation} presents the results of these variants across four datasets using \texttt{Qwen3-8B} and \texttt{GPT-4o-mini}. The results confirm that removing any individual module degrades overall performance. Among all components, the Selection RL-Agent demonstrates the most substantial impact, particularly on complex reasoning tasks, indicating that regulating which peer should be referenced is the primary driver of debate quality. Disabling the Behavior RL-Agent also degrades, and replacing DST fusion with mean-pooling leads to performance decay, demonstrating that mean aggregation fails to distinguish between consensus and divergence among selected peers. Finally, while diverse personas contribute positively, their impact is relatively smaller compared to the RL-Agents. We investigate the performance of various diversity configurations and present the results in Appendix \ref{additional experiments}.

\section{Conclusion and Future Works}
In this paper, we propose DEAR, a novel framework that mitigates blind conformity by dynamically regulating debate relationships from the group perspective. Operating through the What-Who-How stages, DEAR utilizes group evidence to drive two RL-Agents for adaptive peer selection and behavior adjustment. To guarantee coordinated optimization, DEAR employs a multi-agent reinforcement learning algorithm for end-to-end joint training. Extensive experiments demonstrate that DEAR achieves superior performance while significantly reducing token consumption.

In future works, we plan to extend DEAR to more complex scenarios, such as generative tasks and multimodal domains. This exploration will include applications in collaborative coding and open-ended planning, as well as situations involving visual reasoning and cross-modal analysis.

\section{Acknowledgments}
This work was supported by the National Natural Science Foundation of China (Grant No. 62576029) and Beijing Natural Science Foundation (Grant No. L2607023).

{
\small
\bibliographystyle{unsrtnat}
\bibliography{neurips_2026}
}








\appendix
\section{Detailed Related Works}
\label{related_works}
\subsection{Single-Agent Reasoning.}

Large Language Models (LLMs) have shown formidable reasoning capabilities by independently executing complete reasoning processes. However, direct answer generation often fails on complex tasks because it forces models to jump to conclusions without intermediate logical deduction. To address this, some methods \cite{wei2022chain,wang2022self,shum2023automatic,zhao2025spatial,zhao2026stk} have proposed various prompt-based internal reasoning policies. The most representative paradigm is Chain-of-Thought (CoT) \cite{wei2022chain}, which guides models to decompose complex problems into sequential steps through few-shot exemplars or zero-shot instructions such as ``Let's think step by step'' \cite{wei2022chain}. To further enhance robustness and leverage the diversity of generation paths, Self-Consistency Chain-of-Thought (CoT-SC) \cite{wang2022self} samples multiple reasoning trajectories for the same problem and extracts the final answer via majority voting. As research progressed, single-agent reasoning evolved beyond linear paradigms to explore more complex logical topologies. For instance, Tab-CoT \cite{ziqi2023tab} introduces highly structured reasoning. Tree-of-Thoughts (ToT) \cite{yao2023tree} constructs a tree topology by generating multiple candidate answers at each step, facilitating multi-path exploration and self-evaluation. 
Despite these advances, single-agent reasoning remains constrained by the knowledge and reasoning boundaries of an individual model, making it difficult to reliably correct hallucinations and reasoning errors. 

\subsection{Multi-Agent Debate.}
Multi-Agent Debate (MAD) \citep{du2024improving} scales conventional reasoning paradigms by allowing multiple LLMs to engage in multi-round interactions, ultimately reaching a consensus through majority voting. However, fully connected debate incurs substantial token overhead. To improve efficiency, subsequent studies reduce interaction costs by sparsifying debate topologies or eliminating unnecessary debate rounds. S-MAD \citep{li2024improving} and S$^2$-MAD \citep{zeng2025s2} prune redundant information flows, while GD \citep{liu2024groupdebate} restricts global debate to local groups. RUMAD \citep{wang2026rumad} introduces a reinforcement learning-based controller to generate dynamic debate topologies. However, it does not adapt the generation behavior of LLMs. 

Beyond interaction sparsification, several methods further assess response quality. CortexDebate \citep{sun2025cortexdebate} and ConfidenceCal \citep{bai2024confidencecal} use explicit confidence scores to select or weight opinions. MAD-M$^2$ \citep{tian2026multi} filters historical information through agent-generated subjective labels or perplexity. However, confidence, perplexity, and subjective assessments may not reliably reflect reasoning quality, while reliance on internal token probabilities limits applicability to closed-source models.
More importantly, LLMs may prematurely abandon independent judgment under peer influence, resulting in blind conformity and error propagation \citep{yao2025peacemaker}. Free-MAD \citep{cui2025free} mitigates this issue by evaluating individual reasoning trajectories using heuristic scores. Nevertheless, reliably determining reasoning correctness during inference remains difficult without ground-truth supervision.
In contrast, our DEAR framework bypasses flawed individual evaluations by dynamically regulating debate relationships at the group interaction level, effectively protecting the valid reasoning of the minority.

\subsection{The Dempster-Shafer Evidence Theory (DST).}
DST is a theory of belief functions established by Dempster \citep{dempster2008upper}. It combines beliefs from multiple sources through fusion operators, producing a unified belief distribution over the available evidence \citep{shafer1992dempster,josang2012interpretation}. Early studies primarily applied DST to model epistemic uncertainty in classification \citep{fixsen2002modified,denoeux2000neural,tang2023new}. DST has also inspired evidential learning methods that represent predictive beliefs and epistemic uncertainty through Subjective Logic (SL) \citep{jsang2018subjective}. Extending evidential learning to multi-source settings, 
With the development of LLMs, DST has been applied to identify feature-level multimodal conflicts for hallucination detection in large vision-language models (LVLMs) \citep{huang2026detecting,li2025observation,shi2026not,huang2025visual}. In retrieval-augmented generation (RAG), DST is used to resolve conflicts among evidence retrieved from multiple sources \citep{wang2025llm}, while in LLM-as-a-Judge, it supports the fusion of evaluations from multiple agents \citep{liu2024jailjudge}. In this work, we use DST combination rules to fuse the evidence of the LLMs and their selected peers, thereby guiding the adaptation of generation behaviors.


\begin{table*}[t]
\centering
\caption{Performance comparison on \texttt{GPT-4o-mini}. We highlight the \textbf{optimal} and \underline{suboptimal} results. \textbf{Acc. (\%)} denotes accuracy, and \textbf{T. ($K$)} represents the average tokens per task.}
\label{tab:main_results_split_gpt}
\resizebox{\textwidth}{!}{%
\begin{tabular}{cc *{10}{c}}
\toprule

\rowcolor{taskcolor}
\multicolumn{12}{c}{\textbf{Math Reasoning Tasks}} \\

\multirow{3}{*}{\textbf{Architecture}}
& \multirow{3}{*}{\textbf{Method}}
& \multicolumn{2}{c}{\textbf{ID}}
& \multicolumn{6}{c}{\textbf{OOD}}
& \multicolumn{2}{c}{\multirow{2}{*}{\textbf{Avg. (Math)}}} \\
\cmidrule(lr){3-4}
\cmidrule(lr){5-10}

& & \multicolumn{2}{c}{\textbf{GSM8K}}
  & \multicolumn{2}{c}{\textbf{AIME24}}
  & \multicolumn{2}{c}{\textbf{GSM-Hard}}
  & \multicolumn{2}{c}{\textbf{MATH-500}}
  & \multicolumn{2}{c}{} \\

& & Acc. $\uparrow$ & T. $\downarrow$
  & Acc. $\uparrow$ & T. $\downarrow$
  & Acc. $\uparrow$ & T. $\downarrow$
  & Acc. $\uparrow$ & T. $\downarrow$
  & Acc. $\uparrow$ & T. $\downarrow$ \\
\midrule

\multirow{2}{*}{Single-Agent}
& CoT
& 93.0 & 0.5
& \sbest{13.3} & 1.1
& 56.5 & 0.5
& 65.8 & 0.8
& 57.2 & 0.7 \\

& CoT-SC
& \best{96.3} & 6.0
& 10.0 & 17.3
& \sbest{59.3} & 8.5
& \sbest{73.3} & 12.5
& \sbest{59.7} & 11.1 \\
\midrule

\multirow{7}{*}{Multi-Agent}
& Majority Voting
& 94.5 & 1.2
& 6.7 & 3.3
& 57.3 & 1.4
& 63.3 & 2.2
& 55.5 & 2.0 \\

& MAD
& 94.5 & 13.3
& \sbest{13.3} & 36.1
& 54.8 & 15.3
& 64.8 & 24.2
& 56.9 & 22.2 \\

& DMAD
& 93.8 & 17.5
& 10.0 & 37.0
& 55.3 & 19.3
& 72.3 & 27.6
& 57.9 & 25.4 \\

& CortexDebate
& 93.5 & 22.6
& 6.7 & 34.4
& 56.3 & 25.6
& 69.3 & 36.6
& 56.5 & 29.8 \\

& MAD-M$^2$(S)
& 94.0 & 9.0
& 10.0 & 32.3
& 53.5 & 9.7
& 71.0 & 21.1
& 57.1 & 18.0 \\

& ECON
& 91.5 & 8.1
& -- & --
& 52.0 & 9.0
& 65.0 & 12.9
& -- & -- \\

\rowcolor{shadowcolor}
\cellcolor{white}
& DEAR (Ours)
& \sbest{95.0} & 6.3
& \best{16.7} & 18.7
& \best{68.8} & 7.4
& \best{79.8} & 12.0
& \best{65.1} & 11.1 \\

\midrule

\rowcolor{taskcolor}
\multicolumn{12}{c}{\textbf{QA Tasks}} \\

\multirow{3}{*}{\textbf{Architecture}}
& \multirow{3}{*}{\textbf{Method}}
& \multicolumn{2}{c}{\textbf{ID}}
& \multicolumn{6}{c}{\textbf{OOD}}
& \multicolumn{2}{c}{\multirow{2}{*}{\textbf{Avg. (QA)}}} \\
\cmidrule(lr){3-4}
\cmidrule(lr){5-10}

& & \multicolumn{2}{c}{\textbf{ARC-C}}
  & \multicolumn{2}{c}{\textbf{MMLU (Pro.H.)}}
  & \multicolumn{2}{c}{\textbf{TruthfulQA}}
  & \multicolumn{2}{c}{\textbf{GPQA Dia.}}
  & \multicolumn{2}{c}{} \\

& & Acc. $\uparrow$ & T. $\downarrow$
  & Acc. $\uparrow$ & T. $\downarrow$
  & Acc. $\uparrow$ & T. $\downarrow$
  & Acc. $\uparrow$ & T. $\downarrow$
  & Acc. $\uparrow$ & T. $\downarrow$ \\
\midrule

\multirow{2}{*}{Single-Agent}
& CoT
& 91.0 & 0.3
& 71.0 & 0.3
& 68.5 & 0.3
& 38.7 & 1.0
& 67.3 & 0.5 \\

& CoT-SC
& 93.0 & 6.6
& \sbest{73.5} & 7.2
& 70.5 & 7.3
& \sbest{44.0} & 16.2
& \sbest{70.3} & 9.3 \\
\midrule

\multirow{7}{*}{Multi-Agent}
& Majority Voting
& \sbest{93.8} & 1.2
& 73.0 & 1.3
& 68.0 & 1.4
& 40.7 & 3.0
& 68.9 & 1.7 \\

& MAD
& 93.0 & 12.5
& \sbest{73.5} & 19.4
& 72.0 & 14.0
& 40.0 & 32.1
& 69.6 & 19.5 \\

& DMAD
& 92.0 & 17.7
& 71.3 & 25.5
& \sbest{77.8} & 19.2
& 35.3 & 31.8
& 69.1 & 23.6 \\

& CortexDebate
& 92.0 & 12.0
& 70.3 & 15.5
& 74.5 & 11.3
& 42.0 & 42.7
& 69.7 & 20.4 \\

& MAD-M$^2$(S)
& \sbest{93.8} & 6.1
& 73.0 & 10.4
& 62.5 & 7.3
& 42.7 & 20.2
& 68.0 & 11.0 \\

& ECON
& 92.5 & 9.5
& 71.0 & 10.0
& 73.0 & 8.8
& 42.0 & 14.8
& 69.6 & 10.8 \\

\rowcolor{shadowcolor}
\cellcolor{white}
& DEAR (Ours)
& \best{94.5} & 5.6
& \best{74.0} & 8.5
& \best{83.5} & 5.2
& \best{48.0} & 14.3
& \best{75.0} & 8.4 \\

\bottomrule
\end{tabular}%
}
\end{table*}

\section{Experimental Details}
\label{dataset and baseline}
\subsection{Datasets Descriptions.}
Since multi-agent, multi-round debate incurs substantially higher inference costs than standard single-agent evaluation, we randomly sample 400 examples from each benchmark whenever the dataset size permits, using the same sampled subset across all methods for consistent comparison. For open-source models, we use the full AIME24 dataset and randomly sample 150 questions from GPQA Diamond. For closed-source models in the extension experiments, due to API cost constraints, we use 200 examples from GSM-Hard and TruthfulQA for training and 100 non-overlapping examples for evaluation, while using 100 evaluation examples for GPQA Diamond, IMOAnswerBench, and MedXpertQA. The datasets are described below:

\begin{itemize}
    \item \textbf{GSM8K} \citep{cobbe2021training} comprises grade-school mathematics word problems designed to evaluate multi-step arithmetic and basic logical reasoning.

    \item \textbf{AIME24} consists of official problems from the 2024 American Invitational Mathematics Examination and requires rigorous mathematical reasoning.

    \item \textbf{GSM-Hard} \citep{gao2022pal} is a more challenging variant of GSM8K featuring larger and more complex numerical values, placing greater demands on arithmetic reasoning.

    \item \textbf{MATH-500} \citep{lightman2023lets} is curated from the MATH benchmark and covers advanced mathematical domains such as algebra, geometry, and number theory across multiple difficulty levels.

    \item \textbf{ARC-C} \citep{allenai:arc} contains challenging multiple-choice questions from grade-school science examinations and requires scientific knowledge and reasoning beyond superficial text retrieval.

    \item \textbf{MMLU\_Pro (Health)} \citep{wang2024mmlu} is a challenging subset of MMLU-Pro focusing on health and medicine, designed to evaluate professional knowledge and domain-specific reasoning.

    \item \textbf{TruthfulQA} \citep{lin2022truthfulqa} evaluates whether models can provide truthful responses when confronted with common human misconceptions.

    \item \textbf{GPQA Diamond} \citep{rein2024gpqa} is an expert-level benchmark comprising graduate-level questions in physics, biology, and chemistry, requiring specialized scientific knowledge and reasoning.

    \item \textbf{IMOAnswerBench} \citep{luong-etal-2025-towards} contains challenging International Mathematical Olympiad problems and is used to evaluate complex and rigorous mathematical reasoning.

    \item \textbf{MedXpertQA} \citep{zuo2025medxpertqa} focuses on complex clinical medical scenarios. We use its text-modal subset to evaluate domain-specific medical reasoning.
\end{itemize}

\subsection{Baselines Descriptions}
In this section, we provide detailed descriptions of each baseline used in our comparison:
\begin{itemize}
    \item \textbf{CoT (Chain-of-Thought)} \cite{wei2022chain}: Guides large language models to generate intermediate step-by-step reasoning processes before producing the final output, serving as the foundational single-agent prompting method in this study.
    \item \textbf{CoT-SC (Self-Consistency Chain-of-Thought)} \cite{wang2022self}: Builds upon CoT by performing independent sampling to generate multiple distinct reasoning paths (set to 16 in our experiments) and using Majority Voting to determine the final answer.
    \item \textbf{Majority Voting}: Aggregates the initial outputs from multiple independent agents, directly selecting the most frequently occurring answer as the final group decision.
    \item \textbf{MAD (Multi-Agent Debate)} \cite{du2024improving}: Allows multiple agents to read and evaluate each other's reasoning processes across multiple interaction rounds, attempting to reach a group consensus through an iterative feedback mechanism.
    \item \textbf{DMAD (Diverse Multi-Agent Debate)} \cite{liu2025breaking}: Assigns diverse perspectives to agents within the traditional debate framework to mitigate the formation of mental sets during multi-agent interactions.
    \item \textbf{CortexDebate} \cite{sun2025cortexdebate}: Utilizes the internal confidence of the models as a filtering metric to dynamically construct a sparse communication topology, thereby organizing information exchange pathways among agents.
    \item \textbf{MAD-M²} \cite{tian2026multi}: Incorporates subjective (S) and objective (O) methods to filter historical memory. The subjective approach (S) requires agents to autonomously evaluate and label records (e.g., ``YES,'' ``NO,'' or ``NOT SURE''), whereas the objective approach (O) utilizes model perplexity to screen responses.
    \item \textbf{ECON} \cite{yi2025debate}: A multi-agent reinforcement learning-based debate method that strictly freezes the internal parameters of LLMs, training only external networks to adjust and optimize the generation behavior used during LLM API calls.
\end{itemize}

\subsection{Implementation Details}
\label{implementation detail}
Our pre-trained text encoder, $\xi(\cdot)$, employs BGE-M3, which outputs feature embeddings with a dimensionality of 1024. To instantiate the diverse homogeneous debate system introduced in the main text, we assign a distinct persona to each LLM. Specifically, for math reasoning tasks, the designated personas are \texttt{Rigorous\_Formalist}, \texttt{Creative\_Explorer}, and \texttt{Systematic\_Decomposer}; for question-answering (QA) tasks, they are \texttt{Consensus\_Fact\_Checker}, \texttt{Eliminative\_Reasoner}, and \texttt{Careful\_Reader}. To guarantee a fair comparison, all baseline methods are initialized with identical persona configurations. The explicit prompt templates corresponding to each persona are detailed in Appendix \ref{prompt}. The core reinforcement learning hyperparameters and the debate environment parameters adopted during the training process are summarized in Table \ref{tab:hyperparameters}.

\begin{table}[h]
\centering
\caption{Key hyperparameters in the training process of DEAR.}
\label{tab:hyperparameters}
\begin{tabular}{ccc}
\toprule
\textbf{Parameter} & \textbf{Explanation} & \textbf{Value} \\
\midrule

\multicolumn{3}{c}{\textit{HAPPO Optimization Parameters}} \\
\midrule

$\gamma$
& Discount factor
& 1.0 \\

$\lambda$
& GAE parameter
& 0.95 \\

$\epsilon$
& PPO clip range
& 0.2 \\

actor lr
& Actor learning rate
& $5 \times 10^{-4}$ \\

critic lr
& Critic learning rate
& $5 \times 10^{-4}$ \\

$\lambda_v$
& Value loss coefficient
& 0.5 \\

--
& Optimizer type
& Adam \\

num mini-batch
& Number of mini-batches
& 1 \\

ppo epoch
& Number of PPO epochs
& 5 \\

\midrule
\multicolumn{3}{c}{\textit{Debate Environment and Behavior Parameters}} \\
\midrule

$V$
& Number of LLMs
& 3 \\

$T$
& Maximum debate rounds
& 3 \\

$R_{\mathrm{debate}}$
& Reward for correct consensus
& $+1.0$ \\

$T_{\min}/T_{\max}$
& Temperature action boundaries
& $0/1.0$ \\

$P_{\min}/P_{\max}$
& Top-$p$ action boundaries
& $0/1.0$ \\


\bottomrule
\end{tabular}
\end{table}

\section{Analysis for HAPPO with Reward Sharing}

\subsection{Selection of HAPPO}
We consider a cooperative Multi-Agent Debate (MAD) system, where all LLMs collaboratively strive to maximize the accuracy of the group answer. In our proposed DEAR, the Selection RL-Agent and the Behavior RL-Agent form an explicit sequential decision: the Behavior RL-Agent's input is strictly conditioned on the action output of the Selection RL-Agent. Moreover, the two RL-Agents operate over heterogeneous action spaces—the former outputs discrete binary masks, while the latter produces continuous generation behavior. 


Standard algorithms such as Independent PPO (IPPO)
\citep{schulman2017proximal} optimize each RL-Agent independently and do not explicitly account for inter-agent coordination. To coordinate the heterogeneous selection and behavior policies, DEAR adopts the sequential policy-update and importance-weighted advantage propagation mechanism of Heterogeneous-Agent Proximal Policy Optimization (HAPPO) \citep{kuba2021trust}. Specifically, the Selection RL-Agent is updated first, after which its policy importance ratio scales the Behavior RL-Agent's surrogate advantage. This sequential correction accounts for the change in the selection policy when optimizing the conditional behavior policy, providing a principled mechanism for their coordinated optimization.

\subsection{Reward Sharing}
Reward sharing is a standard paradigm in cooperative Multi-Agent Reinforcement Learning (MARL) \cite{yu2022surprising,wu2026think,song2025code,yao2025general,yaorole,song2026highsemanticfidelityhyperdimensional}, premised on the assumption that all agents pursue a unified optimization objective. Our framework DEAR intrinsically satisfies this assumption. The Selection RL-Agent and the Behavior RL-Agent jointly regulate the debate relationships through peer selection and behavior generation to maximize the final accuracy. This collective outcome is inherently inseparable, as it arises from the coupled interplay of peer selection and behavior generation, and cannot be meaningfully attributed to either RL-Agent individually.

Furthermore, reward sharing is not merely a heuristic choice, but a necessity for HAPPO's sequential advantage decomposition to function correctly. The generalized advantage $\hat{A}^t$, derived from a shared reward and a global critic, provides the base signal that is scaled by the Selection RL-Agent's importance sampling ratio to guide the Behavior RL-Agent's update. Without a shared reward, this sequential propagation of policy dependencies would be ill-defined. Therefore, employing reward sharing is both logically sound and theoretically indispensable for the coordinated optimization of our heterogeneous RL-Agents.

\section{Analysis of Computational Efficiency and Training Overhead}
All experiments are conducted on a workstation equipped with an AMD Ryzen 9 9950X 16-Core Processor and a single NVIDIA GeForce RTX 5090 GPU. To accelerate the empirical data collection process, we deploy 16 parallel interaction environments. For open-source models, training DEAR takes approximately 4.6 hours on GSM8K and 4.0 hours on ARC-Challenge. For closed-source models, training takes approximately 3.6 hours on GSM-Hard and 3.2 hours on TruthfulQA. These results demonstrate the training efficiency of DEAR across both open- and closed-source models.

\section{Additional Experiments}
\label{additional experiments}

\subsection{Minority Correction and Conformity Analysis}
\label{appendix:transition_metrics}

This section examines whether DEAR mitigates blind conformity while retaining beneficial correction in multi-agent debate. We focus on two complementary situations: incorrect minority agents adopting a correct majority answer, and correct minority agents conforming to an incorrect majority. We quantify these behaviors using Beneficial Correction Rate (BCR) and Harmful Conformity Rate (HCR), respectively. Both metrics are formally defined below.

\subsubsection{Metric Definitions}
Let the evaluation set contain $N$ problems, each involving $V$ LLMs. We index debate rounds by $t=1,\ldots,T$, where $t=1$ denotes the initial round and $t=T$ denotes the terminal round. $T$ is the total number of debate rounds, and the observed transitions are $t\to t+1$ for $t=1,\ldots,T-1$. For problem $q_n$, let $y_n^\star$ denote the ground-truth answer and $y_{n,i}^{t}$ denote the answer produced by the
$i$-th LLM at round $t$.

When the modal answer is unique, the group majority answer
at round $t$ is defined as:
\begin{equation}
m_n^{t}
=
\operatorname{Mode}
\left(
y_{n,1}^{t},\ldots,y_{n,V}^{t}
\right),
\end{equation}
where $\operatorname{Mode}(\cdot)$ returns the most frequently
occurring answer among the $V$ LLMs.

If two or more answers tie for the highest frequency, no unique
majority answer is defined for that round. We define:
\begin{equation}
\mathcal{U}
=
\left\{
(n,t)\in
\{1,\ldots,N\}\times\{1,\ldots,T-1\}
\;\middle|\;
m_n^t \text{ is uniquely defined}
\right\}
\end{equation}
as the set of problem--round pairs with a unique majority answer
and an observed subsequent round. Since BCR and HCR characterize
minority updates relative to the current majority, they are
computed only over $(n,t)\in\mathcal{U}$.
Terminal-round responses may serve as the destination of the
last transition, but the terminal round itself is excluded from
$\mathcal{U}$ as no subsequent response is generated.

\paragraph{Beneficial Correction Rate (BCR).}
BCR measures the fraction of incorrect minority answers at round
$t$ that change to the correct round-$t$ majority answer at round
$t+1$. For $(n,t)\in\mathcal{U}$, the indicator is defined as:
\begin{equation}
\mathcal{B}_{n,i}^{t}
=
\mathbb{I}\!\left[
y_{n,i}^{t}\neq y_n^\star
\land
y_{n,i}^{t}\neq m_n^{t}
\land
m_n^{t}=y_n^\star
\right].
\end{equation}
BCR is then calculated as:
\begin{equation}
\mathrm{BCR}
=
\frac{
\sum_{(n,t)\in\mathcal{U}}\sum_{i=1}^{V}
\mathcal{B}_{n,i}^{t}
\mathbb{I}\!\left[
y_{n,i}^{t+1}=m_n^{t}
\right]
}{
\sum_{(n,t)\in\mathcal{U}}\sum_{i=1}^{V}
\mathcal{B}_{n,i}^{t}
}.
\end{equation}
A higher BCR indicates more frequent beneficial shifts of
minority answers toward the correct majority.

\paragraph{Harmful Conformity Rate (HCR).}
HCR measures the fraction of correct minority answers at round
$t$ that change to the incorrect round-$t$ majority answer at
round $t+1$. For $(n,t)\in\mathcal{U}$, the indicator is defined as:
\begin{equation}
\mathcal{H}_{n,i}^{t}
=
\mathbb{I}\!\left[
y_{n,i}^{t}=y_n^\star
\land
y_{n,i}^{t}\neq m_n^{t}
\land
m_n^{t}\neq y_n^\star
\right].
\end{equation}
HCR is then calculated as:
\begin{equation}
\mathrm{HCR}
=
\frac{
\sum_{(n,t)\in\mathcal{U}}\sum_{i=1}^{V}
\mathcal{H}_{n,i}^{t}
\mathbb{I}\!\left[
y_{n,i}^{t+1}=m_n^{t}
\right]
}{
\sum_{(n,t)\in\mathcal{U}}\sum_{i=1}^{V}
\mathcal{H}_{n,i}^{t}
}.
\end{equation}
A lower HCR indicates that correct minority answers are less
likely to change to the incorrect majority answer, consistent
with stronger resistance to harmful conformity.

Here, $\mathbb{I}[\cdot]$ denotes the indicator function, which
equals $1$ when the specified condition holds and $0$ otherwise.
BCR reflects beneficial correction and is better when higher,
whereas HCR captures harmful conformity and is better when lower.

\subsubsection{Results and Analysis.}


\begin{table}[htbp]
\centering
\caption{Comparison of debate methods in terms of blind conformity mitigation.
We highlight the \textbf{optimal} and \underline{suboptimal} results.}
\label{tab:transition_analysis}
\resizebox{\textwidth}{!}{%
\begin{tabular}{ccccccccccccc}
\toprule
\multirow{2}{*}{\textbf{Method}}
& \multicolumn{3}{c}{\textbf{GSM8K (ID)}}
& \multicolumn{3}{c}{\textbf{MATH-500 (OOD)}}
& \multicolumn{3}{c}{\textbf{ARC-C (ID)}}
& \multicolumn{3}{c}{\textbf{TruthfulQA (OOD)}} \\
\cmidrule(lr){2-4}
\cmidrule(lr){5-7}
\cmidrule(lr){8-10}
\cmidrule(lr){11-13}
& Acc. $\uparrow$ & HCR $\downarrow$ & BCR $\uparrow$
& Acc. $\uparrow$ & HCR $\downarrow$ & BCR $\uparrow$
& Acc. $\uparrow$ & HCR $\downarrow$ & BCR $\uparrow$
& Acc. $\uparrow$ & HCR $\downarrow$ & BCR $\uparrow$ \\
\midrule

MAD
& \underline{93.5} & 43.3 & 65.2
& 76.3 & 51.4 & \underline{73.9}
& 90.0 & 22.5 & 74.5
& 68.8 & 41.7 & \underline{79.3} \\

DMAD
& 91.5 & 23.8 & 60.6
& \underline{78.8} & 70.8 & 66.4
& 91.5 & 55.0 & 71.1
& 72.0 & 26.7 & 45.5 \\

CortexDebate
& \textbf{94.0} & \underline{19.2} & 69.7
& 69.3 & \underline{23.3} & 59.6
& \underline{91.8} & \underline{16.7} & \underline{83.3}
& 70.8 & \underline{20.8} & 68.3 \\

MAD-M$^2$(O)
& 92.0 & 32.7 & \underline{79.0}
& 77.0 & 70.9 & 69.7
& 90.0 & 62.5 & 79.5
& \underline{73.0} & 31.0 & 78.9 \\

\rowcolor{shadowcolor}
DEAR (Ours)
& \textbf{94.0} & \textbf{12.5} & \textbf{84.2}
& \textbf{85.0} & \textbf{18.3} & \textbf{82.5}
& \textbf{93.3} & \textbf{14.2} & \textbf{94.5}
& \textbf{80.8} & \textbf{18.3} & \textbf{86.7} \\

\bottomrule
\end{tabular}%
}
\end{table}

We conduct these experiments using \texttt{Qwen3-8B} as the
backbone LLM.
As shown in Table \ref{tab:transition_analysis}, DEAR achieves
the highest BCR and lowest HCR across all four datasets,
indicating that incorrect minority agents are more likely to
adopt a correct majority answer, while correct minority agents
are less likely to conform to an incorrect majority.
These results suggest that DEAR's regulation is sensitive to
the evolving group state rather than uniformly encouraging
agreement or independence.
By jointly learning whom to reference and how to generate
subsequent responses, DEAR supports selective revision instead
of indiscriminate alignment with the majority.

\subsection{Analysis of Different Diversity Configurations}
To evaluate the impact of diversity and decouple its benefits from our proposed DEAR, we structure LLM heterogeneity into four progressive levels based on \cite{yang2026understanding}. This approach systematically isolates each diversity source:
\begin{itemize}
    \item L1 (No Diversity): Identical base models and identical default prompts.
    \item L2 (Persona Diversity Only): Identical base models equipped with the persona prompts specified in Section \ref{implementation detail}.
    \item L3 (Model Diversity Only): Distinct base models with identical default prompts.
    \item L4 (Full Diversity): Distinct base models equipped with the persona prompts specified in Section \ref{implementation detail}.
\end{itemize}

In our specific setup, the heterogeneous small-scale LLM group utilizes \texttt{LLaMA-3.1-8B-Instruct} \cite{grattafiori2024llama}, \texttt{GLM-4-9B-0414} \cite{glm2024chatglm}, and \texttt{Qwen3-8B} \cite{qwen3technicalreport}. The heterogeneous large-scale LLM group employs \texttt{DeepSeek-V3.2} \cite{liu2025deepseek}, \texttt{GPT-5.4}, and \texttt{Claude-Sonnet-4.5}. For the homogeneous setting, we use \texttt{Qwen3-8B} and \texttt{Claude-Sonnet-4.5}. 

\begin{table}[htbp]
\centering
\caption{Performance comparison under different diversity configurations for small-scale LLMs.}
\label{tab:diversity_small}
\small
\begin{tabular}{cccccc}
\toprule
\multirow{2}{*}{\textbf{Config.}} & \multirow{2}{*}{\textbf{Method}} & \textbf{GSM8K} & \textbf{GSM-Hard} & \textbf{ARC-C} & \textbf{GPQA Dia.} \\
 & & Acc. (\%) & Acc. (\%) & Acc. (\%) & Acc. (\%) \\
\midrule
 & MAD        & 92.3 & 60.0 & 88.5 & 40.0 \\
\rowcolor{shadowcolor} \cellcolor{white} \multirow{-2}{*}{L1} & DEAR & \textbf{93.3} & \textbf{72.3} & \textbf{92.0} & \textbf{50.7} \\
\midrule
 & MAD    & 93.5 & 61.5 & 90.0 & 40.7 \\
\rowcolor{shadowcolor} \cellcolor{white} \multirow{-2}{*}{L2} & DEAR   & \textbf{94.0} & \textbf{73.5} & \textbf{93.3} & \textbf{52.0} \\
\midrule
 & MAD     & 94.0 & 66.0 & 92.3 & 57.3 \\
\rowcolor{shadowcolor} \cellcolor{white} \multirow{-2}{*}{L3} & DEAR    & \textbf{95.0} & \textbf{77.3} & \textbf{94.5} & \textbf{67.3} \\
\midrule
 & MAD & 94.5 & 66.3 & 93.3 & 62.0 \\
\rowcolor{shadowcolor} \cellcolor{white} \multirow{-2}{*}{L4} & DEAR & \textbf{95.8} & \textbf{78.0} & \textbf{95.5} & \textbf{68.0} \\
\bottomrule
\end{tabular}
\end{table}

\begin{table}[htbp]
\centering
\caption{Performance comparison under different diversity configurations for large-scale LLMs.}
\label{tab:diversity_large}
\small
\begin{tabular}{cccccc}
\toprule
\multirow{2}{*}{\textbf{Config.}} & \multirow{2}{*}{\textbf{Method}} & \textbf{AIME24} & \makecell{\textbf{IMO} \\ \textbf{AnswerBench}} & \makecell{\textbf{MedXpertQA} \\ \textbf{(Text)}} & \textbf{GPQA Dia.} \\
 & & Acc. (\%) & Acc. (\%) & Acc. (\%) & Acc. (\%) \\
\midrule
 & MAD        & 53.3 & 21.0 & 30.0 & 74.0 \\
\rowcolor{shadowcolor} \cellcolor{white} \multirow{-2}{*}{L1} & DEAR & \textbf{70.0} & \textbf{32.0} & \textbf{37.0} & \textbf{80.0} \\
\midrule
 & MAD    & 60.0 & 23.0 & 35.0 & 76.0 \\
\rowcolor{shadowcolor} \cellcolor{white} \multirow{-2}{*}{L2} & DEAR   & \textbf{76.7} & \textbf{35.0} & \textbf{40.0} & \textbf{82.0} \\
\midrule
 & MAD     & 86.7 & 37.0 & 38.0 & 82.0 \\
\rowcolor{shadowcolor} \cellcolor{white} \multirow{-2}{*}{L3} & DEAR    & \textbf{93.3} & \textbf{42.0} & \textbf{42.0} & \textbf{85.0} \\
\midrule
 & MAD & 90.0 & 40.0 & 39.0 & 83.0 \\
\rowcolor{shadowcolor} \cellcolor{white} \multirow{-2}{*}{L4} & DEAR & \textbf{96.7} & \textbf{44.0} & \textbf{45.0} & \textbf{87.0} \\
\bottomrule
\end{tabular}
\end{table}

The results are presented in Tables \ref{tab:diversity_small} and \ref{tab:diversity_large}. Based on this, we draw the following observations:

\textbf{\textit{Observation 1}: Introducing diversity yields progressive performance gains.} From L1 to L4, debate performance improves steadily. Model-level differentiation (L3) serves as the primary driver, while persona prompts (L2/L4) provide only marginal additional gains.

\textbf{\textit{Observation 2}: DEAR consistently outperforms the MAD baseline across all configurations.} In both homogeneous and heterogeneous settings, DEAR surpasses MAD. This indicates that the performance of DEAR stems from the regulation of debate relationships by the two heterogeneous RL-Agents, rather than from a specific diversity configuration.

In summary, diversity configuration and debate relationship regulation are two orthogonal dimensions for improving debate performance. The core innovation of the DEAR lies in the dynamic regulation of debate relationships rather than the introduction of diversity. Moreover, DEAR is compatible with various diversity settings and delivers consistent performance gains under each configuration.

\subsection{Comparison with Sparse Multi-Agent Debate}
To assess DEAR's performance and token efficiency, we compare it with static sparse debate (S-MAD \cite{li2024improving}, S$^2$-MAD \cite{zeng2025s2}) and an RL-based dynamic topology generation method (RUMAD) \cite{wang2026rumad}. Ensuring a fair comparison, we strictly adopt RUMAD’s configuration: a 6-round debate involving 6 LLMs. These LLMs are uniformly instantiated from three base models (LLaMA-3.1-8B-Instruct \cite{grattafiori2024llama}, ChatGLM-4
9B \cite{glm2024chatglm}, and Deepseek-Math-7B-Instruct \cite{guo2025deepseek}), utilizing two agents per model. Performance is assessed on GSM8K \cite{cobbe2021training}, MMLU \cite{hendrycks2020measuring}, and GPQA \cite{rein2024gpqa}. 
\begin{itemize}
    \item MMLU: A comprehensive benchmark evaluating multi-domain reasoning capabilities in LLMs, spanning diverse fields such as STEM, humanities, and social sciences.
    \item GPQA: A highly challenging, graduate-level QA dataset designed to test advanced scientific reasoning across disciplines like physics, biology, and chemistry.
\end{itemize}

\begin{table}[htbp]
\centering
\caption{Comparison of accuracy and token consumption with sparse debate methods. \textbf{Acc. (\%)} denotes accuracy, and \textbf{T. ($K$)} represents the average tokens per task. S-MAD$_{*}$ and S-MAD$_o$ represent S-MAD configured with star and ring topologies, respectively. Best results are \textbf{bolded} and the second-best are \underline{underlined}.}
\label{tab:sparse_results}
\begin{tabular}{ccccccccc}
\toprule
\multirow{2}{*}{\textbf{Method}} & \multicolumn{2}{c}{\textbf{GSM8K}} & \multicolumn{2}{c}{\textbf{MMLU}} & \multicolumn{2}{c}{\textbf{GPQA}} & \multicolumn{2}{c}{\textbf{Avg.}} \\
\cmidrule(lr){2-3} \cmidrule(lr){4-5} \cmidrule(lr){6-7} \cmidrule(lr){8-9}
& Acc. $\uparrow$ & T. $\downarrow$ & Acc. $\uparrow$ & T. $\downarrow$ & Acc. $\uparrow$ & T. $\downarrow$ & Acc. $\uparrow$ & T. $\downarrow$ \\
\midrule
S-MAD$_{*}$ & 83.0 & 39.4 & 61.0 & 33.3 & 30.0 & 36.4 & 58.0 & 36.4 \\
S-MAD$_o$  & 70.0 & 37.8 & 54.0 & 31.7 & 34.0 & 38.6 & 52.7 & 36.0 \\
S$^2$-MAD  & 70.0 & 30.5 & 46.0 & 25.4 & 28.0 & \textbf{23.3} & 48.0 & 26.4 \\
RUMAD      & \underline{89.0} & \underline{17.3} & \underline{68.0} & \underline{22.5} & \underline{35.0} & 33.4 & \underline{64.0} & \underline{24.4} \\
\rowcolor{shadowcolor} DEAR      & \textbf{94.0} & \textbf{16.7} & \textbf{74.0} & \textbf{21.8} & \textbf{42.0} & \underline{31.3} & \textbf{70.0} & \textbf{23.3} \\
\bottomrule
\end{tabular}
\end{table}

Table \ref{tab:sparse_results} show that static sparse methods (S-MAD and S$^2$-MAD) reduce token costs by restricting communication, yet typically sacrifice accuracy. RUMAD improves this accuracy-cost balance via RL-driven dynamic topologies, though it solely optimizes peer selection. DEAR outperforms these methods by delivering higher accuracy at lower token costs across all benchmarks. This confirms that the synergistic regulation of both Selection and Behavior RL-Agents offers a far more efficient performance boost than static or single-dimensional dynamic topologies.

\subsection{Analyzing Cross-Domain Zero-Shot Transfer}
The task-agnostic of DEAR stems from deriving RL-Agents inputs from objective cosine similarity. Beyond validating cross-task generalization within single domains in our primary experiments, we further conduct zero-shot cross-domain transfer experiments (Math $\leftrightarrow$ QA).
Specifically, for the Math $\rightarrow$ QA transfer, we train the RL-Agents on GSM8K and evaluate their zero-shot performance on TruthfulQA and GPQA Diamond. Conversely, in the QA $\rightarrow$ Math transfer, the RL-Agents are trained on ARC-C and tested on unseen reasoning tasks, including GSM-Hard and MATH-500.

\begin{table}[htbp]
\centering
\caption{Cross-domain transfer performance of DEAR between Math and QA on \texttt{GPT-4o-mini}. Best results are \textbf{bolded} and the second-best are \underline{underlined}.}
\label{tab:cross_domain_results}
\small
\begin{tabular}{cccc}
\toprule
\rowcolor{taskcolor} \multicolumn{4}{c}{\textbf{Math $\rightarrow$ QA}} \\
\textbf{Method} & \textbf{GSM8K} & \textbf{TruthfulQA} & \textbf{GPQA Diamond} \\
 & Acc. (\%) & Acc. (\%) & Acc. (\%) \\
\midrule
MAD & \underline{94.5} & 72.0 & 40.0 \\
DEAR (in domain) & \textbf{95.0} & \textbf{83.5} & \textbf{48.0} \\
\rowcolor{shadowcolor} DEAR (out of domain) & - & \underline{76.0} & \underline{45.0} \\
\midrule
\rowcolor{taskcolor} \multicolumn{4}{c}{\textbf{QA $\rightarrow$ Math}} \\
\textbf{Method} & \textbf{ARC-C} & \textbf{GSM-Hard} & \textbf{MATH-500} \\
 & Acc. (\%) & Acc. (\%) & Acc. (\%) \\
\midrule
MAD & \underline{93.0} & 54.8 & 64.8 \\
DEAR (in domain) & \textbf{94.5} & \textbf{68.8} & \textbf{79.8} \\
\rowcolor{shadowcolor} DEAR (out of domain) & - & \underline{64.5} & \underline{76.0} \\
\bottomrule
\end{tabular}
\end{table}

Table \ref{tab:cross_domain_results} demonstrates that DEAR outperforms the standard MAD across all target datasets, proving that the RL-Agents rely on task-agnostic features to form a universal regulation policy. The slight performance decay in cross-domain transfer is primarily caused by the natural distribution shift of the input features. Even with a constant cosine similarity, the fundamental differences in the statistical distributions of these scores across domains (e.g., Math vs. QA) perturb the RL-Agents' inputs, leading to suboptimal decision-making.

\subsection{Ablation on SL and DST Fusion}
To validate the effectiveness of Subjective Logic (SL) and the DST fusion in debate relationship regulation, we design two ablation variants: (1) \textbf{DEAR-EviMean}, which directly feeds group evidence into the Selection RL-Agent and applies mean-pooling for evidence fusion; and (2) \textbf{DEAR-EviDST}, which retains the same input but employs Dempster-Shafer Evidence Theory (DST) to fuse the received evidence. The experiments are conducted using \texttt{GPT-4o-mini}.
\begin{table}[htbp]
\centering
\caption{Performance comparison of DEAR with different regulation bases and fusion strategies.}
\label{tab:ablation_study}
\begin{tabular}{c c c c c}
\toprule
\rowcolor{taskcolor} 
& GSM8K & GSM-Hard & ARC-C & GPQA Diamond \\
Method & Acc. (\%) & Acc. (\%) & Acc. (\%) & Acc. (\%) \\
\midrule
DEAR-EviMean & 92.0 & 64.0 & 91.0 & 42.0 \\
DEAR-EviDST & \underline{93.3} & \underline{65.5} & \underline{92.3} & \underline{42.7} \\
\rowcolor{shadowcolor} 
DEAR & \textbf{95.0} & \textbf{68.8} & \textbf{94.5} & \textbf{48.0} \\
\bottomrule
\end{tabular}
\end{table}

The results in Table \ref{tab:ablation_study} indicate performance degradation in both variants. Notably, DEAR-EviDST underperforms the full DEAR framework despite both using the DST mechanism. This gap proves the effectiveness of identifying LLMs' consultation tendencies and quantifying individual consultation uncertainty. Consequently, they provide a reliable basis for group-level debate regulation. Furthermore, the steeper decline in DEAR-EviMean shows that DST fuses conflicting evidence more rationally than simple mean-pooling.

\section{Case Study}
\label{sec:case_study}

To intuitively demonstrate how DEAR mitigates blind conformity in MAD,
we extract a real debate trajectory from the challenging GPQA Diamond \citep{rein2024gpqa}. In addition to the LLMs' responses,
we report the LLM-specific sampling parameters $T$ and
$Top\_p$ selected by the behavior policy at each debate state.

\begin{promptbox}{Example of Debate Trajectory in GPQA Diamond}
\begin{itemize}
    \item \textbf{Initial question $\mathcal{Q}$}:
    \begin{itemize}
        \item[-] Your colleague has devised a new quantum field theory on four-dimensional spacetime, and is exploring the regularization of certain higher-order loop diagrams in that theory. On their desk you spy a scribbled note: a Feynman diagram, and next to it, the words ``size estimate'' followed by a string of physical constants, numbers, and ratios between what appear to be energy scales. The symbols read: $\alpha^3 \cdot g^2 \sqrt{2} \cdot 8 \cdot \frac{1}{(4\pi)^6} \cdot \left(\frac{Q}{M}\right)^2$. The diagram, unfortunately, has been rendered unreadable by a spilled coffee. How many loops did it contain? \\
        (A) 1 \quad (B) 2 \quad (C) 6 \quad (D) 3
    \end{itemize}

    \item \textbf{Ground Truth}: (D) 3

    \item \textbf{Initial Sampling Parameters}: \\
    All three LLMs use the same temperature initialization, while
    retaining LLM-specific $Top\_p$ values:
    \[
        T^{(0)}=[0.00,\,0.00,\,0.00],
        \qquad
        Top\_p^{(0)}=[0.82,\,0.88,\,0.87].
    \]

    \item \textbf{Initial Round}:
    \begin{itemize}
        \item[-] \textbf{[Agent 1] Persona: Consensus\_Fact\_Checker} \\
        \textit{Reasoning process:} In quantum electrodynamics (QED), the size estimate of a Feynman diagram typically scales with $\alpha^n$, where $n$ is the number of loops. In this case, the expression includes $\alpha^3$, suggesting a three-loop diagram. \\
        \textit{Conclusion:} The most consistent interpretation with well-established quantum field theory is a three-loop diagram. \texttt{\textbackslash boxed\{D\}} \cmark

        \item[-] \textbf{[Agent 2] Persona: Eliminative\_Reasoner} \\
        \textit{Reasoning process:} The presence of a factor of $1/(4\pi)^6$ is a strong indicator of six-loop diagrams. Evaluating options: (A) 1 loop gives $1/(4\pi)^2$. (B) 2 loops give $1/(4\pi)^4$. (C) 6 loops matches the $1/(4\pi)^6$ factor. (D) 3 loops would give $1/(4\pi)^6$ only in non-standard schemes. \\
        \textit{Conclusion:} By elimination, 6 is the best-supported choice. \texttt{\textbackslash boxed\{C\}} \xmark

        \item[-] \textbf{[Agent 3] Persona: Careful\_Reader} \\
        \textit{Reasoning process:} In dimensional regularization, the divergence is proportional to $1/(4\pi)^{d-4}$. For four-dimensional regularization, we get a factor of $1/(4\pi)^6$, which implies six loops. \\
        \textit{Conclusion:} The factor $1/(4\pi)^6$ strongly suggests six loops. \texttt{\textbackslash boxed\{C\}} \xmark
    \end{itemize}

    \item \textbf{Sampling Parameters for Round 1}: \\
    After peer selection, the Behavior RL-Agent produces
    the following parameters from the fused evidence:
    \[
        T^{(1)}=[0.10,\,0.75,\,0.70],
        \qquad
        Top\_p^{(1)}=[0.75,\,0.98,\,0.97].
    \]

    \item \textbf{Round 1}:
    \begin{itemize}
        \item[-] \textbf{[Agent 1] Persona: Consensus\_Fact\_Checker} \textit{($\leftarrow$ Isolated)} \\
        \textit{Reasoning process:} Independently re-checking the reasoning. In quantum field theory, particularly in dimensional regularization, the divergence of a Feynman diagram is often expressed with a factor of the form $1/(4\pi)^{2n}$, where $n$ is the number of loops. Given the factor $1/(4\pi)^6$, we deduce that $n=3$, since $(4\pi)^{2 \cdot 3} = (4\pi)^6$. \\
        \textit{Conclusion:} The presence of $\alpha^3$ also supports this. Therefore, the diagram contains three loops. \texttt{\textbackslash boxed\{D\}} \cmark

        \item[-] \textbf{[Agent 2] Persona: Eliminative\_Reasoner} \textit{($\leftarrow$ Refers to Agents 1 \& 3)} \\
        \textit{Reasoning process:} Scanning peers (Agent 1, Agent 3). Agent 1 misinterprets the scaling of the coupling constants, suggesting $\alpha^3$ implies three loops. Agent 3 correctly identifies the $1/(4\pi)^6$ factor as a strong indicator of six loops based on the exponent mapping. \\
        \textit{Conclusion:} While $1/(4\pi)^6$ could be interpreted as three loops in some contexts, the standard convention corresponds to six loops. By elimination, 6 remains the logical choice. \texttt{\textbackslash boxed\{C\}} \xmark

\item[-] \textbf{[Agent 3] Persona: Careful\_Reader}
\textit{($\leftarrow$ Refers to Agents 1 \& 2)} \\
\textit{Reasoning process:} Scanning peers (Agent 1, Agent 2).
Agent 1 favors three loops based on the power of $\alpha$,
while Agent 2 interprets the exponent of $(4\pi)$ as the loop
count. Neither argument alone establishes the correct mapping.
Re-checking the size estimate, the usual four-dimensional
loop-counting factor is $1/(16\pi^2)$ per loop. Thus an
$L$-loop contribution carries a factor
$(16\pi^2)^{-L}=(4\pi)^{-2L}$. Matching the given factor
$(4\pi)^{-6}$ gives $2L=6$, hence $L=3$.
My initial answer incorrectly equated the exponent of
$(4\pi)$ directly with the number of loops. \\
\textit{Conclusion:} Under the usual four-dimensional
loop-counting convention, the diagram contains three loops.
\texttt{\textbackslash boxed\{D\}} \cmark
    \end{itemize}

    \item \textbf{Sampling Parameters for Round 2}: \\
    After Round 1, the behavior policy accordingly updates the sampling parameters to
    \[
        T^{(2)}=[0.10,\,0.80,\,0.10],
        \qquad
        Top\_p^{(2)}=[0.65,\,0.99,\,0.74].
    \]

    \item \textbf{Round 2}:
    \begin{itemize}
        \item[-] \textbf{[Agent 1] Persona: Consensus\_Fact\_Checker} \textit{($\leftarrow$ Refers to Agent 2)} \\
        \textit{Reasoning process:} Scanning peer (Agent 2). Agent 2 still argues that the $1/(4\pi)^6$ factor is a strong indicator of six-loop diagrams. However, independently confirming the derivation, the standard convention is that the exponent is $2n$. Thus $2n=6 \Rightarrow n=3$. \\
        \textit{Conclusion:} The peer’s reasoning remains flawed. Mathematical derivation clearly indicates three inherent loops. \texttt{\textbackslash boxed\{D\}} \cmark
        
        \item[-] \textbf{[Agent 2] Persona: Eliminative\_Reasoner} \textit{($\leftarrow$ Refers to Agents 1 \& 3)} \\
        \textit{Reasoning process:} Scanning peers (Agent 1, Agent 3). Both peers now point out the specific dimensional regularization formula $1/(4\pi)^{2L}$. This means the exponent is twice the number of loops, so $2L=6 \Rightarrow L=3$. Let's systematically eliminate: (A) 1 gives $1/(4\pi)^2$. (B) 2 gives $1/(4\pi)^4$. (C) 6 would correspond to $1/(4\pi)^{12}$. (D) 3 matches the $1/(4\pi)^6$ factor. \\
        \textit{Conclusion:} By eliminating the incorrect exponent mappings based on the proper $2L$ rule, 3 loops is the only defensible answer. \texttt{\textbackslash boxed\{D\}} \cmark

\item[-] \textbf{[Agent 3] Persona: Careful\_Reader}
\textit{($\leftarrow$ Refers to Agent 2)} \\
\textit{Reasoning process:} Scanning peer (Agent 2).
Agent 2's previous-round response still identifies six loops
by equating the exponent of $(4\pi)$ with the loop count.
However, the usual four-dimensional loop factor is
$1/(16\pi^2)=(4\pi)^{-2}$ per loop. Therefore,
$(4\pi)^{-6}$ corresponds to three loops; six loops would
instead give $(4\pi)^{-12}$. The peer's response provides
no valid reason to revise my corrected answer. \\
\textit{Conclusion:} The loop-factor counting continues
to support three loops.
\texttt{\textbackslash boxed\{D\}} \cmark
    \end{itemize}

    \item \textbf{Final Vote}:
    \begin{itemize}
        \item[-] \texttt{\textbackslash boxed\{D\}}. \cmark
    \end{itemize}
\end{itemize}
\end{promptbox}
This trajectory illustrates that the Selection RL-Agent adaptively regulates debate relationships by controlling which peers should be referenced, while the Behavior RL-Agent dynamically adjusts \(T\) and \(Top\_p\) based on the fused evidence from selected peers, rather than uniformly increasing sampling stochasticity across all LLMs.

\section{Prompt Set}
\label{prompt}
In our framework, each LLM is assigned a distinct persona prompt. Each persona defines a unique reasoning style, guiding the LLM to approach problems from a differentiated perspective. The detailed persona prompts are presented below.

\subsection{Math Reasoning Tasks}

For math reasoning tasks, we assign the following three personas from \cite{yang2026understanding}:

\begin{promptbox}{Persona 1: Rigorous\_Formalist}
You are a rigorous mathematician who formalizes problems precisely. 

Your approach:

- Define all variables and terms clearly at the start.

- State any assumptions explicitly.

- Justify each step with mathematical principles or rules.

- Use precise mathematical notation and language.

- Ensure logical completeness in your reasoning.
\end{promptbox}

\begin{promptbox}{Persona 2: Creative\_Explorer}
You are an innovative problem solver who looks for elegant and creative solutions.

Your approach:

- Look for patterns, shortcuts, and elegant solutions.

- Try multiple approaches and compare them.

- Think about the problem from different angles.

- Don't be afraid to try unconventional methods.

- Value insight and elegance alongside correctness.
\end{promptbox}

\begin{promptbox}{Persona 3: Systematic\_Decomposer}

You are an expert at breaking complex problems into manageable parts. 

Your approach:

- Identify the core components of the problem.

- Break the problem into smaller, independent sub-problems.

- Solve each sub-problem systematically.

- Carefully combine the results, checking for consistency.

- Review the overall solution for completeness.
\end{promptbox}

\subsection{Question Answering Tasks}

For QA tasks, we assign the following three personas from \cite{yang2026understanding}:

\begin{promptbox}{Persona 1: Consensus\_Fact\_Checker}

You select the option most consistent with well-established consensus knowledge.

Your approach:

- Prefer mainstream scientific/medical/historical understanding when relevant.

- Penalize claims that conflict with widely accepted facts.

- Choose the option whose core claim best matches reliable background knowledge.
\end{promptbox}

\begin{promptbox}{Persona 2: Eliminative\_Reasoner}

You are an expert at process of elimination reasoning.

Your approach:

- Systematically evaluate each option.

- Find clear reasons to eliminate wrong answers.

- Use contradictions and impossibilities.

- Narrow down to the most defensible answer.

- Verify the remaining answer makes sense.
\end{promptbox}

\begin{promptbox}{Persona 3: Careful\_Reader}

You are a careful reader who avoids traps in questions.

Your approach:

- Pay attention to qualifiers (most likely, best, depends, always/never).

- Identify what the question is REALLY asking (definition vs application).

- Prefer simple textbook truths over tricky interpretations.
\end{promptbox}

\subsection{Debate Prompt Templates}

In addition to persona prompts, we design prompt templates for the initial round and subsequent debate rounds.

\subsubsection{Initial Round Prompt Template}

In the initial round, each LLM independently solves the problem without peer interaction:

\begin{promptbox}{Initial Round Prompt}
\{Persona Prompt\}

Problem:
\{Question\}

Instruction:

1. Provide your reasoning process.

2. Summarize your reasoning into a conclusion that includes the final answer.

\{Output Instruction\}
\end{promptbox}

\noindent Where the output instructions are:
\begin{itemize}
    \item \textbf{Math}: ``Output the final result inside \texttt{\textbackslash boxed\{\}}. Example: \texttt{\textbackslash boxed\{42\}}.''
    \item \textbf{QA}: ``Output the final choice selection using the format: \texttt{\textbackslash boxed\{X\}} where X is the exact option letter. Example: \texttt{\textbackslash boxed\{C\}}.''
\end{itemize}

\subsubsection{Debate Round Prompt Template}

In subsequent debate rounds, each LLM receives its own previous response along with peers' responses. Depending on whether peer responses are available, two prompt variants are used.

\textbf{Case 1: With Peer Responses}
\begin{promptbox}{Debate Round Prompt (With Peer Responses)}
\{Persona Prompt\}

Problem:
\{Question\}

Previous Thought:
\{Self Memory\}

Peer Responses:
\{Peer Context\}

REVIEW PROTOCOL:

1. BEFORE reading peers, rederive the key steps independently.

2. THEN scan peers. Only consider an objection if it identifies a SPECIFIC step where a particular number, formula, or factual claim is wrong.

Instruction:

1. Provide your reasoning process.

2. Summarize your reasoning into a conclusion that includes the final answer.

\{Output Instruction\}
\end{promptbox}

\textbf{Case 2: Without Peer Responses}

Note that when no peer responses are available, the peer context block is replaced with:

\begin{promptbox}{Debate Round Prompt (No Peer Responses Available)}
\{Persona Prompt\}

Problem:
\{Question\}

Previous Thought:
\{Self Memory\}

Task: Independently re-check the reasoning. Identify the single step
least certain, re-derive it from scratch, and correct it only if a concrete error is found.

Instruction:

1. Provide your reasoning process.

2. Summarize your reasoning into a conclusion that includes the final answer.

\{Output Instruction\}

\end{promptbox}

\section{The Pseudocode of Training Process of DEAR}
\label{pseudocode}
The training pseudocode for DEAR is presented in Algorithm \ref{alg:dear}.

\begin{algorithm}[htbp]
  \caption{DEAR}
  \label{alg:dear}
  \textbf{Input:} LLMs $\mathcal{V}=\{1,\dots,V\}$, Debate rounds $T$, Text encoder $\xi(\cdot)$, Problem $\mathcal{Q}$ \\
  \textbf{Initialize:} Selection RL-Agent $\rho_\phi$, Behavior RL-Agent $\pi_\theta$, Critic $V_\psi$, Replay buffer $\mathcal{D}$ \\
  \textbf{Set:} learning rate $lr$, hyperparameters $\gamma, \lambda, \epsilon$
  \begin{algorithmic}[1]
  \FOR{episode $= 1, \dots, E$}
      
      \FOR{round $t = 1$ to $T$}
          \STATE \textbf{1. Multi-Agent Debate Execution:}
          \FOR{each LLM $i \in \mathcal{V}$}
              \STATE Generate response $Ans_i^t$ based on problem $\mathcal{Q}$ with parameters $(T_i^t, Top\_p_i^t)$
          \ENDFOR
          
          \STATE \textbf{2. Group Evidence Extraction and Decouple (What):}
          \STATE Extract group evidence $e_{ij}$ from pairwise responses via $\xi(\cdot)$
          \STATE Decouple $e_{ij}$ into belief mass $b_{ij}$ and uncertainty $u_i$ via Subjective Logic
          \STATE Construct group observation $o^{\rho}$ from $\{b_i, u_i, ID_i\}_{i=1}^{V}$
          
          \STATE \textbf{3. Adaptive Debater Selection (Who) \& Generation Behavior Adjustment (How):}
          \STATE Sample peer selection mask $\boldsymbol{ids}^t \sim \rho_\phi(\cdot \mid o^{\rho})$ via independent Bernoulli distributions
          
          \FOR{each LLM $i \in \mathcal{V}$}
              \STATE Filter reference peers $\mathcal{X}_i^t = \{j \mid ids_{i \leftarrow j}^t = 1\}$
              \STATE Construct next debate context using problem $\mathcal{Q}$ and peer responses $\{Ans_j^t \mid j \in \mathcal{X}_i^t\}$
              
              \STATE Fuse evidence of LLM $i$ with selected peers $\mathcal{X}_i^t$ by DST combination rule
              \STATE Derive fused evidence $\hat{E}_i$ via inverse SL mapping
              \STATE Sample continuous latent variable $z_i^{t} \sim \pi_\theta(\cdot \mid \hat{E}_i)$ from diagonal Gaussian
              \STATE Map $z_i^{t}$ to generation behavior $a_i^{t+1} = [T_i^{t+1}, Top\_p_i^{t+1}]$ via Sigmoid
          \ENDFOR
      \ENDFOR
      
       \STATE Obtain terminal reward $R_{debate}$ based on final consensus; Store trajectory in $\mathcal{D}$
      \STATE Sample a minibatch of trajectories from $\mathcal{D}$
      
      \STATE \textbf{4. End-to-End Joint Optimization:}
      \STATE Compute Generalized Advantage Estimation (GAE) $\hat{A}^t$ 
      \STATE Compute sequential surrogate advantage $M^{\pi,t} = r_{\phi}^{\rho,t}\hat{A}^t$
      \STATE Update Selection RL-Agent $\rho_\phi$ by minimizing objective $\mathcal{L}_{\rho}$ (Eq. \ref{loss:selection})
      \STATE Update Behavior RL-Agent $\pi_\theta$ by minimizing objective $\mathcal{L}_{\pi}$ (Eq. \ref{loss:behavior})
      \STATE Update Critic $V_\psi$ by minimizing objective $\mathcal{L}_{critic}$ (Eq. \ref{loss:critic})
  \ENDFOR
  \RETURN Trained $\rho_\phi$ and $\pi_\theta$
  \end{algorithmic}
\end{algorithm}

\clearpage
\newpage
\input{checklist.tex}

\end{document}

%% file: checklist.tex
\section*{NeurIPS Paper Checklist}

\begin{enumerate}

\item {\bf Claims}
    \item[] Question: Do the main claims made in the abstract and introduction accurately reflect the paper's contributions and scope?
    \item[] Answer: \answerYes{} 
    \item[] Justification: See Abstract and Introduction.

\item {\bf Limitations}
    \item[] Question: Does the paper discuss the limitations of the work performed by the authors?
    \item[] Answer: \answerYes{} 
    \item[] Justification: See Conclusion and Future Works.

\item {\bf Theory assumptions and proofs}
    \item[] Question: For each theoretical result, does the paper provide the full set of assumptions and a complete (and correct) proof?
    \item[] Answer: \answerYes{} 
    \item[] Justification: See Appendix.

    \item {\bf Experimental result reproducibility}
    \item[] Question: Does the paper fully disclose all the information needed to reproduce the main experimental results of the paper to the extent that it affects the main claims and/or conclusions of the paper (regardless of whether the code and data are provided or not)?
    \item[] Answer: \answerYes{} 
    \item[] Justification: See Appendix.

\item {\bf Open access to data and code}
    \item[] Question: Does the paper provide open access to the data and code, with sufficient instructions to faithfully reproduce the main experimental results, as described in supplemental material?
    \item[] Answer: \answerYes{} 
    \item[] Justification: See supplemental material.

\item {\bf Experimental setting/details}
    \item[] Question: Does the paper specify all the training and test details (e.g., data splits, hyperparameters, how they were chosen, type of optimizer) necessary to understand the results?
    \item[] Answer: \answerYes{} 
    \item[] Justification: See Appendix.

\item {\bf Experiment statistical significance}
    \item[] Question: Does the paper report error bars suitably and correctly defined or other appropriate information about the statistical significance of the experiments?
    \item[] Answer: \answerYes{} 
    \item[] Justification: See Experiments.

\item {\bf Experiments compute resources}
    \item[] Question: For each experiment, does the paper provide sufficient information on the computer resources (type of compute workers, memory, time of execution) needed to reproduce the experiments?
    \item[] Answer: \answerYes{} 
    \item[] Justification: See Appendix.
    
\item {\bf Code of ethics}
    \item[] Question: Does the research conducted in the paper conform, in every respect, with the NeurIPS Code of Ethics \url{https://neurips.cc/public/EthicsGuidelines}?
    \item[] Answer: \answerYes{} 
    \item[] Justification: This paper does not involve human subjects, sensitive personal data, or applications intended to cause harm.

\item {\bf Broader impacts}
    \item[] Question: Does the paper discuss both potential positive societal impacts and negative societal impacts of the work performed?
    \item[] Answer: \answerYes{} 
    \item[] Justification: See Conclusion and Future Works.
    
\item {\bf Safeguards}
    \item[] Question: Does the paper describe safeguards that have been put in place for responsible release of data or models that have a high risk for misuse (e.g., pre-trained language models, image generators, or scraped datasets)?
    \item[] Answer: \answerNA{} 
    \item[] Justification: Do not involve.

\item {\bf Licenses for existing assets}
    \item[] Question: Are the creators or original owners of assets (e.g., code, data, models), used in the paper, properly credited and are the license and terms of use explicitly mentioned and properly respected?
    \item[] Answer: \answerYes{} 
    \item[] Justification: See Appendix.

\item {\bf New assets}
    \item[] Question: Are new assets introduced in the paper well documented and is the documentation provided alongside the assets?
    \item[] Answer: \answerNA{} 
    \item[] Justification: Do not involve.

\item {\bf Crowdsourcing and research with human subjects}
    \item[] Question: For crowdsourcing experiments and research with human subjects, does the paper include the full text of instructions given to participants and screenshots, if applicable, as well as details about compensation (if any)? 
    \item[] Answer: \answerNA{} 
    \item[] Justification: Do not involve.

\item {\bf Institutional review board (IRB) approvals or equivalent for research with human subjects}
    \item[] Question: Does the paper describe potential risks incurred by study participants, whether such risks were disclosed to the subjects, and whether Institutional Review Board (IRB) approvals (or an equivalent approval/review based on the requirements of your country or institution) were obtained?
    \item[] Answer: \answerNA{} 
    \item[] Justification: Do not involve.

\item {\bf Declaration of LLM usage}
    \item[] Question: Does the paper describe the usage of LLMs if it is an important, original, or non-standard component of the core methods in this research? Note that if the LLM is used only for writing, editing, or formatting purposes and does \emph{not} impact the core methodology, scientific rigor, or originality of the research, declaration is not required.
    \item[] Answer: \answerYes{} 
    \item[] Justification: The core methodology explicitly focuses on Multi-Agent Debate systems, where LLMs serve as the primary interacting agents to mitigate blind conformity. 

\end{enumerate}